\documentclass[final,5p,times,twocolumn]{elsarticle}

\usepackage[pscoord]{eso-pic}
\usepackage{graphicx}
\usepackage{amssymb}
\usepackage{lineno}
\usepackage{hyperref}

\journal{Nuclear Instruments and Methods in Physics Research A }

\begin{document}

\begin{frontmatter}


%%\title{Title\tnoteref{label1}}
%%\tnotetext[label1]{text}
%%\fntext[label3]{aaa}

\title{MCNP6 fragmentation of light nuclei at intermediate energies}

%% \author[label1,label2]{Andrei Seryi}

\author[label1]{Stepan G. Mashnik}   \ead{mashnik@lanl.gov}
\author[label1,label2]{Leslie M. Kerby}
%% \author[label6]{Add Missing Authors}
\address[label1]{Los Alamos National Laboratory, Los Alamos, NM 87545, USA}
\address[label2]{University of Idaho, Moscow, ID 83844, USA}
%% \address[label6]{Add Missing Authors}

%\begin{center}
\address{\sf \large In memory of Dick Prael, outstanding scientist and person}
%\end{center}

%%\author{Andrei Seryi}
%%\address{JAI, Oxford}

\begin{abstract}
Fragmentation reactions induced on light target nuclei
by protons and light nuclei of energies 
around 1 GeV/nucleon and below  are studied with 
the latest Los Alamos Monte Carlo transport code MCNP6 and with its 
cascade-exciton model (CEM) and Los Alamos version of the quark-gluon 
string model (LAQGSM) event generators, version 03.03, used as stand-alone 
codes. Such reactions are involved in different applications, like 
cosmic-ray-induced single event upsets (SEU's), radiation protection, and 
cancer therapy with proton and ion beams, 
among others;
therefore, 
it is important that MCNP6 simulates them as well as possible. CEM and 
LAQGSM assume that intermediate-energy fragmentation reactions on light 
nuclei occur generally in two stages. The first stage is the intranuclear 
cascade (INC), followed by the second, Fermi breakup disintegration of 
light excited residual nuclei produced after INC. Both CEM and LAQGSM 
account also for coalescence of light fragments (complex particles) up 
to $^4$He from energetic nucleons emitted during INC. We investigate the 
validity and performance of MCNP6, CEM, and LAQGSM in simulating fragmentation 
reactions at intermediate energies and discuss possible ways of further 
improving these codes.
\end{abstract}

\begin{keyword}
Monte Carlo
\sep transport codes
\sep MCNP6
\sep cascade-exciton model (CEM)
\sep Los Alamos version of the quark-gluon string model (LAQGSM)
\sep fragmentation
\sep Fermi breakup
\sep coalescence
\sep fragment spectra
\sep production cross-sections


\end{keyword}

\end{frontmatter}

% \linenumbers


\section{Introduction}
\label{sec-intro}



Fragmentation reactions induced by protons and light nuclei of energies 
around 1 GeV/nucleon and below on light target nuclei are involved in 
different applications, like cosmic-ray-induced single event upsets 
(SEU's), radiation protection, and cancer therapy with proton and ion 
beams, 
among others.
It is impossible to measure all nuclear 
data needed for such applications; therefore, Monte Carlo transport 
codes are usually used to simulate impacts associated with fragmentation 
reactions. It is important that available transport codes simulate such 
reactions as well as possible. For this reason, during the past several 
years, efforts have been done to investigate the validity and performance 
of, and to improve where possible, nuclear reaction models 
simulating fragmentation of light nuclei in GEANT4 \cite{1},
%[1], 
SHIELD-HIT 
\cite{2}--\cite{4},
%[2-4], 
and PHITS \cite{5, 6}.
%[5, 6].

The Los Alamos Monte Carlo transport code MCNP6 [7] uses the latest 
version of the cascade-exciton model (CEM) as incorporated in its event 
generator CEM03.03 [8, 9] to simulate fragmentation of light nuclei at 
intermediate energies for reactions induced by nucleons, pions, and 
photons, and the Los Alamos version of the quark-gluon string model 
(LAQGSM) as implemented in the code LAQGSM03.03 [9, 10] to simulate 
fragmentation reactions induced by nuclei and by particles at higher 
energies, up to about 1 TeV/nucleon.

In recent years, MCNP6, with its CEM and LAQGSM event generators,
has been
extensively validated and verified (V\&V) against 
a large variety of nuclear reactions on both thin and thick targets 
(see, e.g. Refs. [11-14] and references therein), but was never 
tested
specifically 
on fragmentation of light nuclei at intermediate energies.
To address
this, we 
investigate the performance of MCNP6, CEM, and LAQGSM in 
simulating fragmentation reactions at intermediate energies and discuss 
possible ways of further improving these codes.

\section{A Brief Survey of CEM and LAQGSM Physics}
\label{sec-cem-laq}

Details, examples of results, and useful references to different versions 
of CEM and LAQGSM may be found in a recent lecture \cite{9}.
%[9].

The Cascade-Exciton Model (CEM) of nuclear reactions was proposed more 
than 30 years ago at the Laboratory of Theoretical Physics, JINR, Dubna, 
USSR by Gudima, Mashnik, and Toneev \cite{15}.
%[15]. 
It is based on the standard (non time-dependent) Dubna IntraNuclear 
Cascade (INC) \cite{16, 17}
%[16, 17] 
and the Modified Exciton Model (MEM) \cite{18, 19}.
%[18, 19]. 
The code LAQGSM03.03 is the latest modification \cite{10}
%[10] 
of LAQGSM \cite{20},
%[20], 
which in its turn is an improvement of the Quark-Gluon String Model 
(QGSM) \cite{21}.
%[21]. 
It describes reactions induced by both particles and nuclei at incident 
energies up to about 1 TeV/nucleon.

The basic version of both the CEM and LAQGSM event generators is the 
so-called ``03.03'' version, namely CEM03.03 \cite{8, 9, 22}
%[22] 
and LAQGSM03.03 \cite{9, 10, 23}.
%[23]. 
The CEM code calculates nuclear reactions induced by nucleons, pions, 
and photons. It assumes that the reactions occur generally in three stages 
(see Fig.~1). The first stage is the INC, in which primary particles can be 
re-scattered and produce secondary particles several times prior to absorption 
by, or escape from the nucleus. When the cascade
stage of a reaction is completed, CEM uses the coalescence model to 
``create'' high-energy d, t, $^3$He, and $^4$He by final-state interactions 
among emitted cascade nucleons outside of the target. The emission 
of the cascade particles determines the particle-hole configuration, $Z$, $A$, 
and the excitation energy that is the starting point for the second, 
preequilibrium stage of the reaction. The subsequent relaxation of the nuclear 
excitation is treated in terms of an improved version of the modified exciton 
model of preequilibrium decay followed by the equilibrium evaporation/fission 
stage. 

%\begin{figure*}[htb!]
\begin{figure}[h!]
%\begin{figure*}[h!]
\centering
\includegraphics[width=0.48\textwidth]{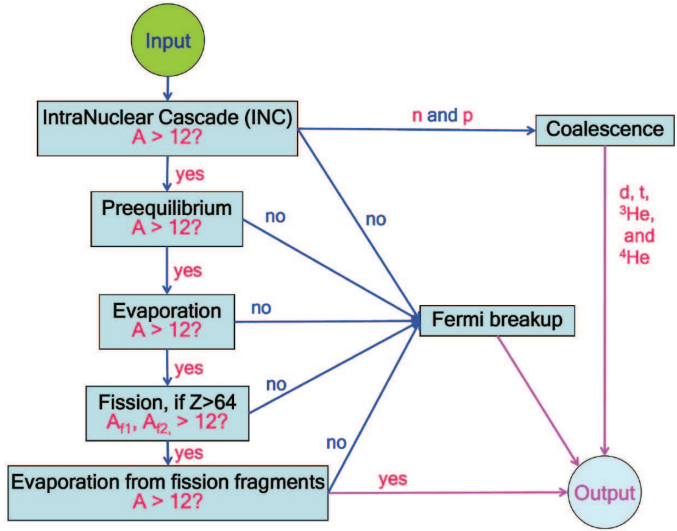}
%\includegraphics[width=0.48\textwidth]{Untitled-bw.eps}
%\vspace*{3mm}
\caption{Flow chart of nuclear-reaction calculations by CEM03.03 and LAQGSM03.03.}
\label{fig:1}
%\end{figure*}
\end{figure}

Generally, all three components may contribute to 
experimentally measured particle spectra and other distributions. But if the 
residual nuclei after the INC have atomic numbers with $A \leq A_{Fermi} = 12$, 
CEM uses the Fermi breakup model to calculate their further disintegration 
instead of using the preequilibrium and evaporation models. Fermi breakup is 
much faster to calculate and gives results very similar to the continuation 
of the more detailed models to much lighter nuclei. LAQGSM also 
describes nuclear reactions, generally, as a three-stage process: INC, 
followed by preequilibrium emission of particles during the equilibration 
of the excited residual nuclei formed after the INC, followed by evaporation 
of particles from or fission of the compound nuclei. LAQGSM was developed 
with a primary focus on describing reactions induced by nuclei, as well as 
induced by most elementary particles, at high energies, up to about 
1 TeV/nucleon. The INC of LAQGSM is completely different from the one in 
CEM. LAQGSM also considers Fermi breakup of nuclei with $A \leq 12$ 
produced after the cascade, and the coalescence model to ``produce'' 
high-energy d, t, $^3$He, and $^4$He from nucleons emitted during the INC.

Many people participated in the CEM and LAQGSM code development over their 
more than 40-year history. Current contributors to their ``03.03'' versions 
are S. G. Mashnik, K. K. Gudima, A. J. Sierk, R. E. Prael, M. I. Baznat, and 
N. V. Mokhov. One of the authors (L.M.K.) has jointed these efforts recently to 
extend the preequilibrium models of CEM and LAQGSM by accounting for possible 
emission of light fragment (LF) heavier than $^4$He, up to $^{28}$Mg.

\subsection{The Intranuclear Cascade Mechanism}
\label{sec-2.1}

The INC approach is based on the ideas of Heisenberg and Serber, who 
regarded intranuclear cascades as a series of successive quasi-free 
collisions of the fast primary particle with the individual nucleons 
of the nucleus. Basic assumptions of and conditions for INC applicability 
may be found in \cite{9}.
%[9]. 
Comprehensive details and
useful references are published in \cite{16, 17}.
%[16, 17].

\subsubsection{The INC of CEM03.03}
\label{sec-2.1.1}

The intranuclear cascade model in CEM03.03 is based on the standard 
(non-time-dependent) version of the Dubna cascade model 
\cite{16, 17}.
%[16, 17]. 
All the cascade calculations are carried out in a three-dimensional 
geometry. The nuclear matter density 
$\rho(r)$
is described by a Fermi distribution with two parameters taken from 
the analysis of electron-nucleus scattering. For simplicity, the 
target nucleus is divided by concentric spheres into seven zones in 
which the nuclear density is considered to be constant. The energy 
spectrum of the target nucleons is estimated in the perfect Fermi-gas 
approximation. The influence of intranuclear nucleons on the incoming 
projectile is taken into account by adding to its laboratory kinetic 
energy an effective real potential, as well as by considering the Pauli 
principle which forbids a number of intranuclear collisions and 
effectively increases the mean free path of cascade particles inside 
the target. The interaction of the incident particle with the nucleus 
is approximated as a series of successive quasi-free collisions of the 
fast cascade particles ($N$, $\pi$, or $\gamma$) with intranuclear 
nucleons.

The integral cross sections for the free $N N$, $\pi N$, and $\gamma N$ 
interactions are approximated in the Dubna INC model 
\cite{16, 17}
%[16, 17] 
using a special algorithm of interpolation/extrapolation through a 
number of picked points, mapping as well as possible the experimental 
data. This was done very accurately by Prof. Barashenkov's group
using all experimental data available at that time, more than 45 years 
ago 
\cite{24}.
%[24]. 
Currently the experimental data on cross sections is much more complete 
than at that time; therefore we have revised the approximations of all 
the integral elementary cross sections used in CEM.

The kinematics of two-body elementary interactions and absorption of 
photons and pions by a pair of nucleons is completely defined by a 
given direction of emission of one of the secondary particles. The 
cosine of the angle of emission of secondary particles in the c.m. 
system is calculated by the Dubna INC with approximations based on 
available experimental data. For elementary interactions with more 
than two particles in the final state, the Dubna INC uses the statistical 
model to simulate the angles and energies of products (see details in 
\cite{16}).
%[16]).

For the improved version of the INC in CEM03.03, we use currently available 
experimental data and recently published systematics proposed by other 
authors and have developed new approximations for angular and energy 
distributions of particles produced in nucleon-nucleon and photon-proton 
interactions. 
In addition, we have incorporated into newer versions of CEM 
a possibility to normalize the final results to systematics based on available 
experimental reaction cross sections. The condition for the transition from 
the INC stage of a reaction to preequilibrium was changed; on the whole, 
the INC stage in CEM03.03 is longer while the preequilibrium stage is shorter 
in comparison with previous versions. We have incorporated real binding 
energies for nucleons in the cascade instead of the approximation of a 
constant separation energy of 7 MeV used in the initial versions of the CEM 
and have imposed momentum-energy conservation for each simulated even 
(conservation was only
 ``on the average'' in earlier versions). 
Along with the improved elementary cross sections, we
 also changed and improved 
the algorithms of many INC routines and many INC routines were rewritten, 
which  significantly speeded up the code.
Details, examples of results, and references to this portion of our work 
may be found in 
\cite{9}.
%[9].

\subsubsection{The INC of LAQGSM03.03}
\label{sec-2.1.2}

The INC of LAQGSM03.03 is described with a recently improved version 
\cite{10, 23, 25}
%[10, 23, 25] 
of the time-dependent intranuclear cascade model developed initially at JINR 
in Dubna, often referred to in the literature as the Dubna intranuclear 
Cascade Model, DCM (see 
\cite{26}
%[26] 
and references therein). The DCM models interactions of fast cascade particles 
(``participants'') with nucleon spectators of both the target and projectile 
nuclei and includes as well interactions of two participants (cascade particles). 
It uses experimental cross sections at energies below 4.5 GeV/nucleon, and those 
calculated by the Quark-Gluon String Model 
\cite{21, 27}
%[21, 27] 
at higher energies to simulate angular and energy distributions of cascade 
particles, and also considers the Pauli Exclusion Principle.

In contrast to the CEM version of the INC described above, DCM uses 
a continuous nuclear density distribution; therefore, it does not need to consider 
refraction and reflection of cascade particles inside or on the border 
of a nucleus. It also keeps track of the time of an intranuclear collision 
and of the depletion of the nuclear density during the development of the 
cascade (the so-called ``trawling effect'') and takes into account the 
hadron formation time.

All the new approximations developed recently for the INC of CEM 
to describe total cross sections and elementary energy and angular 
distributions of secondary particles from hadron-hadron interactions 
have been incorporated also into the INC of LAQGSM 
\cite{23}.
%[23]. 
In addition, a new high-energy photonuclear reaction model based on the 
event generators for $\gamma p$ and $\gamma n$ reactions from the Moscow INC 
\cite{28}
%[28] 
(kindly provided to us by Dr. Igor Pshenichnov) and on the latest photonuclear 
version of CEM 
\cite{29}
%[29] 
was developed and incorporated into the INC of LAQGSM; this allows us 
to calculate reactions induced by photons with energies of up to tens of 
GeV. In the latest version of LAQGSM
\cite{10},
%[10], 
the INC was modified for a better description of nuclear reactions at very 
high energies (above 20 GeV/nucleon). Finally, the algorithms of many 
LAQGSM INC routines were revised and some INC routines were rewritten, which 
speeded up the code significantly.
Details, examples of results, and 
references to this portion of our work may be found in 
\cite{9}.
%[9].

\subsection{The Coalescence Model}
\label{sec-2.2}

When the cascade stage of a reaction is completed, CEM and LAQGSM
use the coalescence model described in Ref. 
\cite{26}
%[26] 
to ``create'' high-energy d, t, $^3$He, and $^4$He by final-state interactions 
among emitted cascade nucleons outside of the target nucleus. In 
contrast to most other coalescence models for heavy-ion-induced reactions, 
where complex-particle spectra are estimated simply by convolving the measured 
or calculated inclusive spectra of nucleons with corresponding fitted 
coefficients, CEM03.03 and LAQGSM03.03 use in their simulations of particle 
coalescence real information about all emitted cascade nucleons and do not 
use integrated
spectra. We assume that all the cascade nucleons having differences in their 
momenta smaller than $p_c$ and the correct isotopic content form an appropriate 
composite particle. The coalescence radii $p_c$ were fitted for each composite 
particle in Ref. 
\cite{26}
%[26] 
to describe available data for the reaction Ne+U at 1.04 GeV/nucleon, but the 
fitted values turned out to be quite universal and were subsequently found to 
describe high-energy complex-particle production satisfactorily for a variety 
of reactions induced both by particles and nuclei at incident energies up to 
about 200 GeV/nucleon, when describing nuclear reactions with different 
versions of LAQGSM 
\cite{9}
%[9] 
or with its predecessor, the Quark-Gluon String Model (QGSM) 
\cite{21}.
%[21]. 
These parameters are:
\begin{eqnarray}
p_c(d) & = & 90 \mbox{ MeV/c ;} \nonumber \\
p_c(t) & = & p_c(^3{\mbox He}) = 108 \mbox{ MeV/c ;} \\
p_c(^4{\mbox He}) & = & 115 \mbox{ MeV/c .} \nonumber 
\label{e1}
\end{eqnarray}

As the INC of CEM is different from those of LAQGSM or 
QGSM, it is natural to expect different best values for $p_c$ 
as well. Our recent studies show that the values of parameters 
$p_c$ defined by Eq. (1) are also good for CEM for projectile 
particles with kinetic energies $T_0$ lower than 300 MeV and equal to 
or above 1 GeV. For incident energies in the interval 
300 MeV $ \leq T_0 < 1$ GeV, a better overall agreement with the 
available experimental data is obtained by using values of $p_c$ 
equal to 150, 175, and 175 MeV/c for d, t ($^3$He), and $^4$He, 
respectively. These values of $p_c$ are fixed as defaults in 
 CEM03.03. If several cascade nucleons are 
chosen to coalesce into composite particles, they are removed from 
the distributions of nucleons and do not contribute further to 
such nucleon characteristics as spectra, multiplicities, {\it etc}.

In comparison with the initial version 
\cite{26},
%[26], 
in CEM03.03 and LAQGSM03.03, several coalescence routines have been 
changed/deleted and have been tested against a large variety of measured 
data on nucleon- and nucleus-induced reactions at different incident 
energies.

\subsection{Preequilibrium Reactions}
\label{sec-2.3}

The subsequent preequilibrium interaction stage of nuclear reactions 
is considered by our current CEM and LAQGSM in the framework of the 
latest version of the Modified Exciton Model (MEM) 
\cite{18, 19}
%[18, 19] 
as described in Ref.
\cite{22}.
%[22]. 
At the preequilibrium stage of a reaction, we take into account all 
possible nuclear transitions changing the number of excitons $n$ with 
$\Delta = +2$, -2, and 0, as well as all possible multiple subsequent 
emissions of n, p, d, t, $^3$He, and $^4$He. The corresponding system 
of master equations describing the behavior of a nucleus at the 
preequilibrium stage is solved by the Monte-Carlo technique~\cite{15}.
%[15].

CEM considers the possibility of fast d, t, $^3$He, and $^4$He emission 
at the preequilibrium stage of a reaction in addition to the emission 
of nucleons. We assume that in the course of a reaction $p_j$ excited 
nucleons (excitons) are able to condense with probability $\gamma_j$ 
forming a complex particle which can be emitted during the preequilibrium 
state. The ``condensation'' probability $\gamma_j$ is estimated as the 
overlap integral of the wave function of independent nucleons with that 
of the complex particle (see details in~\cite{15})
%[15])

\begin{equation}
\gamma_j \simeq p^3_j (V_j / V)^{p_j - 1} =  p^3_j (p_j / A)^{p_j - 1}
\mbox{ .}
\label{e2}
\end{equation}

This is a rather crude estimate. 
As is frequently done,
 the values $\gamma_j$ 
are taken from fitting the theoretical preequilibrium spectra to the 
experimental ones. In CEM, to improve the description of 
preequilibrium complex-particle emission, we estimate $\gamma_j$ by 
multiplying the estimate provided by Eq. (2) by an empirical coefficient 
$M_j (A,Z,T_0)$ whose values are fitted to available nucleon-induced 
experimental complex-particle spectra. 

CEM and LAQGSM predict forward-peaked (in the laboratory system) 
angular distributions for preequilibrium particles. For instance, 
CEM assumes that a nuclear state with a given excitation energy 
$E^*$ should be specified not only by the exciton number $n$ but also 
by the momentum direction $\bf \Omega$. This calculation scheme is easily 
realized by the Monte-Carlo technique 
\cite{15}.
%[15]. 
It provides a good description of double differential spectra of preequilibrium 
nucleons and a not-so-good but still satisfactory description of 
complex-particle spectra from different types of nuclear reactions at incident 
energies from tens of MeV to several GeV. For incident energies below about 
200 MeV, Kalbach 
\cite{30}
%[30] 
has developed a phenomenological systematics for preequilibrium-particle 
angular distributions by fitting available measured spectra of nucleons 
and complex particles. As the Kalbach systematics are based on measured 
spectra, they describe very well the double-differential spectra of 
preequilibrium particles and generally provide a better agreement of 
calculated preequilibrium complex-particle spectra with data than does 
the CEM approach 
\cite{15}.
%[15]. 
This is why we have incorporated into CEM03.03 and LAQGSM03.03 the 
Kalbach systematics 
\cite{30}
%[30] 
to describe angular distributions of both preequilibrium nucleons and complex 
particles at incident energies up to 210 MeV. At higher energies, we use the 
CEM approach 
\cite{15}.
%[15].

The standard version of the CEM [15] provides an overestimation 
of preequilibrium particle emission from different reactions we have 
analyzed (see more details in 
\cite{31}).
%[31]). 
One way to solve this problem, suggested in Ref. 
\cite{31},
%[31] 
is to change the criterion for the transition from the cascade stage 
to the preequilibrium one. Another easy way, suggested in Ref. 
\cite{31},
%[31] 
to shorten the preequilibrium stage of a reaction is to arbitrarily 
allow only transitions that increase the number of excitons, 
$\Delta n = +2$, {\it i.e.}, only allow the evolution of a nucleus 
toward the compound nucleus. In this case, the time of the 
equilibration will be shorter and fewer preequilibrium particles 
will be emitted, leaving more excitation energy for the evaporation. 
This approach was used in the CEM2k 
\cite{31}
%[31] 
version of the CEM and it allowed us to describe much better the p+A 
reactions measured at GSI in inverse kinematics at energies around 
1 GeV/nucleon. Nevertheless, the ``never-come-back'' approach seems 
unphysical; therefore we no longer use it. We now address the problem 
of emitting fewer preequilibrium particles in the CEM by following 
Veselsky 
\cite{32}.
%[32]. 
We assume that the ratio of the number of quasi-particles (excitons) 
$n$ at each preequilibrium reaction stage to the number of excitons 
in the equilibrium configuration $n_{eq}$, corresponding to the same 
excitation energy, to be a crucial parameter for determining the 
probability of preequilibrium emission $P_{pre}$ (see details in 
\cite{9, 22, 32}).
%[9, 22, 32]).
Algorithms of many preequilibrium routines were changed and almost all 
these routines were rewritten, which has speeded up the code significantly
relative to earlier versions
\cite{9, 22}.
%[9, 22].

\subsection{Evaporation}
\label{sec-2.4}

CEM and LAQGSM use an extension of 
the Generalized Evaporation Model (GEM) code GEM2 by Furihata 
\cite{33}
%[33] 
after the preequilibrium stage of reactions to describe evaporation of 
nucleons, complex particles, and light fragments heavier than $^4$He 
(up to $^{28}$Mg) from excited compound nuclei and to describe 
%their 
fission, if the compound nuclei are heavy enough to fission 
($Z \geq 65$).

When including evaporation of up to 66 types of particles 
in GEM2, running times increase significantly compared to the 
case when evaporating only 6 types of particles, up to $^4$He. 
The major particles emitted from an excited nucleus are n, p, d, 
t, $^3$He, and $^4$He. For most cases, the total emission probability 
of particles heavier than $\alpha$ is negligible compared to those for 
the emission of light ejectiles. Our detailed investigation of different 
reactions shows that if we study only nucleon and complex-particle 
spectra or only spallation and fission products and are not interested 
in light fragments, we can consider evaporation of only 6 types of 
particles in GEM2 and save much time, getting results very close to 
the ones calculated with the more time consuming ``66'' option. 
In
our current code versions, we allow the number of types of evaporated 
particles to be
selected in advance.
A 
detailed description of GEM2, as incorporated into CEM and LAQGSM, may 
be found in 
\cite{9, 22}.
%[9, 22].

\subsection{Fission}
\label{sec-2.5}

The fission model used in GEM2 is based on Atchison's model 
\cite{34},
%[34], 
often referred in the literature as the Rutherford Appleton Laboratory 
(RAL) fission model, which is where Atchison developed it. The mass-, 
charge-, and kinetic energy-distribution of fission fragments are 
simulated by RAL using approximations based on available experimental 
data (see details in 
\cite{9, 22, 33, 34}).
%[9, 22, 33, 34]). 
For CEM03.03 and LAQGSM03.03, we modified slightly 
\cite{35}
%[35] 
GEM2. Since in 
this study we consider only reactions on light,
not fissioning nuclei, we will not discuss further the fission model; 
 interested readers may found details and 
further references in 
\cite{9, 22, 35, 36}.
%[9, 22, 35, 36].

\subsection{The Fermi Breakup Model}
\label{sec-2.6}

After calculating the coalescence stage of a reaction, CEM and 
LAQGSM move to the description of the last slow stages of the 
interaction, namely to preequilibrium decay and evaporation, with a 
possible competition of fission. But 
at any stage,
if the residual 
nuclei have atomic numbers with $A \leq A_{Fermi} = 12$, CEM and LAQGSM use 
the Fermi breakup model 
\cite{37}
%[37] 
to calculate their further disintegration instead of using the 
preequilibrium and evaporation models.  
All formulas and details of the algorithms 
used in the version of the Fermi breakup model developed in the group 
of the Late Prof. Barashenkov at 
JINR, Dubna, 
may be found in 
\cite{38}; we use this model.
%[38].

The original version of the model contained a few features which very occasionally
could lead to unphysical fragments;
these
could cause problems in a transport model. All
these issues have been dealt with in the current version, which no longer 
encounters
such
problems.

\section{Results}
\label{sec-3}

As described 
above, de-excitation of light nuclei with
$A \leq A_{Fermi}$
produced after the INC
is described by CEM and LAQGSM only with the Fermi break-up model, where
$A_{Fermi}$ is a ``cut-off value'' fixed in our models.
The value of $A_{Fermi}$ is a model 
dependent parameter, not a physics characteristic of nuclear reactions. 
Actually, the initial version of the Fermi breakup model we incorporated 
in CEM and LAQGSM 
\cite{22, 23}
%[22, 23] 
was used when $A \leq A_{Fermi} = 16$, just as $A_{Fermi} = 16$ is used 
currently in GEANT4 (see 
\cite{1})
%[1]) 
and in SHIELD-HIT (see 
\cite{2}--\cite{4}).
%[2-4]). 
But as mentioned in Section 2.6, that initial version of the Fermi breakup 
model had some problems and crashed our codes in some cases. To avoid 
unphysical results and code crashes, we chose 
the expedient of using
 $A_{Fermi} = 12$ in 
both CEM and LAQGSM. Later, 
we fixed the problems in the Fermi break-up model, but did not at that time change the
value of
$A_{Fermi}$, and never studied how its value affects the final 
results calculated 
in these
codes. We address this here, calculating spectra of emitted particles and light fragments,
and yields of all possible products from various reactions using different values for
$A_{Fermi}$. 
We discuss 
below separately product cross sections (Section 3.1) and spectra of 
particles and light fragments (Section 3.2).

\subsection{Fragment production cross sections}
\label{sec-3.1}

One of the most difficult tasks for any theoretical model is to predict 
cross sections of arbitrary products as functions of the incident energy 
of the projectiles initiating the reactions, {\it i.e.}, excitation functions. 
Therefore, we chose to start our study with comparing the available 
experimental data on excitation functions of products from several 
proton-induced reactions on light nuclei at intermediate energies 
with predictions by MCNP6 using its default event generator for such 
reactions, CEM03.03, as well as with results calculated by CEM03.03 
used as a stand-alone code.

Figs. 2--15 present examples of excitation functions for all products 
we found at least several measured values for proton-induced reactions on 
$^{14}$N, $^{16}$O, $^{27}$Al, and $^{28}$Si. To understand better the 
reasons of agreements or disagreements of calculated values with the 
measured excitation functions, we present in our figures also the total 
reaction cross sections, experimental and theoretical.

Figs. 2--5 show our results for the p + $^{14}$N reaction. The first 
thing 
%we like 
to note is that the total reaction cross sections simulated 
with MCNP6 and shown in the upper-left plot in Fig. 2 with small solid 
circles agree well with the available experimental data (symbols) 
and with calculations by CEM03.03 used as a stand-alone code (solid
line).
There is a difference between the models,
  especially in the regions of incident proton 
energies $T_p = 50 - 100$ MeV and $T_p \ge 2$ GeV. 
To be expected, since
 MCNP6 and CEM03.03 use very similar, but slightly different 
approximations for the total proton-nucleus reaction cross sections 
(see details and references in 
\cite{7, 8}).
%[7, 8]). 
These little differences in the total reaction cross sections will produce, 
respectively, similar differences in all excitation functions 
simulated with MCNP6 and CEM03.03.

The total reaction cross sections are based on systematics (see details 
and references in 
\cite{7, 8}),
%[7, 8]), 
therefore 
they
do not depend on the value of $A_{Fermi}$ we use in our 
calculations. However, we performed calculations of all excitation 
functions shown in Figs. 2 to 5 with CEM03.03 used as a stand-alone code 
with its ``default value'' $A_{Fermi} = 12$, as well as with a modification 
of the code using $A_{Fermi} = 16$, which in case of these p + $^{14}$N reactions, 
actually corresponds to $A_{Fermi} = 14$.
We cannot get a mass number A = 16 
from p + $^{14}$N interactions, and even a nucleus with $A = 15$ would not be 
produced by the INC of CEM03.03 at these intermediate energies.

First, from the results presented in Figs. 2 to 5, we see a very good agreement 
between the excitation functions simulated by MCNP6 using CEM03.03 and 
calculations by CEM03.03 used as a stand-alone code, and a reasonable agreement 
with most of available experimental data. This fact serves as a validation 
and verification (V\&V) of MCNP6 and shows no problems with the implementation 
of CEM03.03 in MCNP6 or with the simulations of these reactions by either code.

Second, we'd like to  explicitly inform the readers that we do not worry too
much about some observed discrepancies between some calculated excitation 
functions and measured data at low energies, below 20 MeV. As the default, MCNP6 
uses data libraries at
such low energies and never uses CEM03.03 or its  other event generators, if 
data libraries are available (MCNP6 has proton-induced data libraries for the 
reactions studied here). 
By contrast, CEM uses its INC to simulate the first stage of nuclear reactions, and the
INC is not supposed to work properly at such low energies
(see details in 
\cite{8, 9}).
%[8, 9]).

Third, results calculated both with
 $A_{Fermi} = 12$ and 16 
agree 

\begin{figure*}[htb!]
\centering
\includegraphics[width=1.0\textwidth]{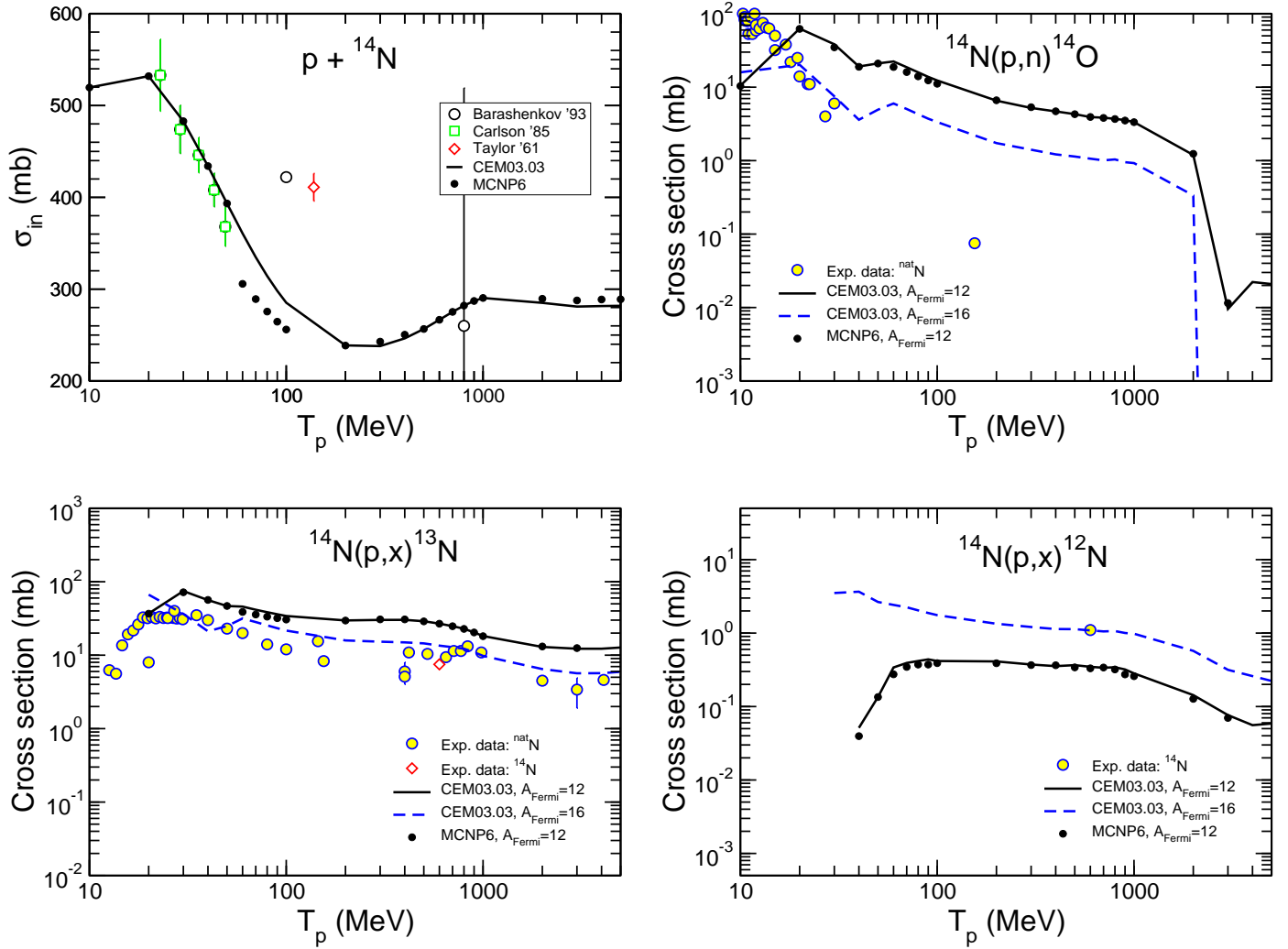}
\caption{Total inelastic cross section and excitation functions for the production 
of $^{14}$O, $^{13}$N, and $^{12}$N from 
p + $^{14}$N calculated with CEM03.03 using the ``standard'' version of the Fermi 
breakup model ($A_{Fermi} = 12$) and with a cut-off value of 
16 for $A_{Fermi}$, as well as with MCNP6 using CEM03.03 
($A_{Fermi} = 12$) compared with experimental data,
as indicated. 
Experimental data for inelastic cross sections are from Refs. 
\cite{39}--\cite{41},
%[39-41], 
while the 
data for excitation functions are from the T16 Lib compilation 
\cite{42}.
%[42].
}
\label{fig:2}
\end{figure*}
{\noindent
reasonably well with available data, taking into account that all 
calculations, at all energies and for all reactions were done with the fixed 
version of our codes, without any tuning or changing of any parameters. However, 
in some cases, we can observe significant differences between excitation 
functions calculated with $A_{Fermi} = 12$ and 16.}

For this particular reaction, the excitation functions for the production of 
$^{14}$O, $^{13}$N, $^{12}$N, $^{13}$C, $^{12}$C, and $^{10}$C calculated with 
$A_{Fermi} = 16$ (that for our p + $^{14}$N reaction is the same as 
$A_{Fermi} = 14$, which from a physical point of view means that we use only 
Fermi breakup after INC and never use preequilibrium and/or evaporation models 
to calculate this reaction) agree better with available experimental data 
than results obtained with $A_{Fermi} = 12$. On the other hand, excitation 
functions for the production of $^9$Be and $^7$Be are reproduced better with 
$A_{Fermi} = 12$.

Figs. 6 to 9 present results similar to the ones shown in Figs. 2 to 5, but 
for the reaction p + $^{16}$O.
Most of the experimental data for these reactions were measured
on $^{nat}$O, with only a few data points obtained for $^{16}$O; all our 
calculations were performed for $^{16}$O. For these reactions, we performed 
three sets of calculations, using $A_{Fermi} = 12$, 14, and 16 in CEM03.03. 
The general agreement/disagreement of our results with available measured 
data for oxygen is very similar to what we showed above for p + $^{14}$N, 
with 
the major difference that almost all products from oxygen are
better predicted with
 $A_{Fermi} = 14$; 
production of $^{11}$B is described a little better with $A_{Fermi} = 16$, 
while $^9$Be and $^7$Be are reproduced better with $A_{Fermi} = 12$, just 
as for nitrogen (see Figs. 4 and 5).

Figs. 10 to 15 show results similar to those
 in Figs. 2--9, 
but for proton interactions with $^{27}$Al and $^{28}$Si.
All reactions 
on silicon were calculated for $^{28}$Si, while most of the data were 
measured from $^{nat}$Si (see details in legends of Fig. 14). Aluminum 
and silicon are interesting 
%to us 
because they are used in many applications. 
From a theoretical point of view, p + $^{27}$Al and $^{28}$Si reactions are 
challenging because Al and Si are relatively light, with significant 
contributions from 
the Fermi breakup models in our simulations. At 
the same time Al and Si have mass numbers higher than the discussed above,
 allowing some significant
contribution 
to the calculated values from preequilibrium and evaporation 
processes. On the whole, the agreement of the results
 with 
available measured data for Al and Si is very similar

\begin{figure*}[htb!]
\centering
\includegraphics[width=1.0\textwidth]{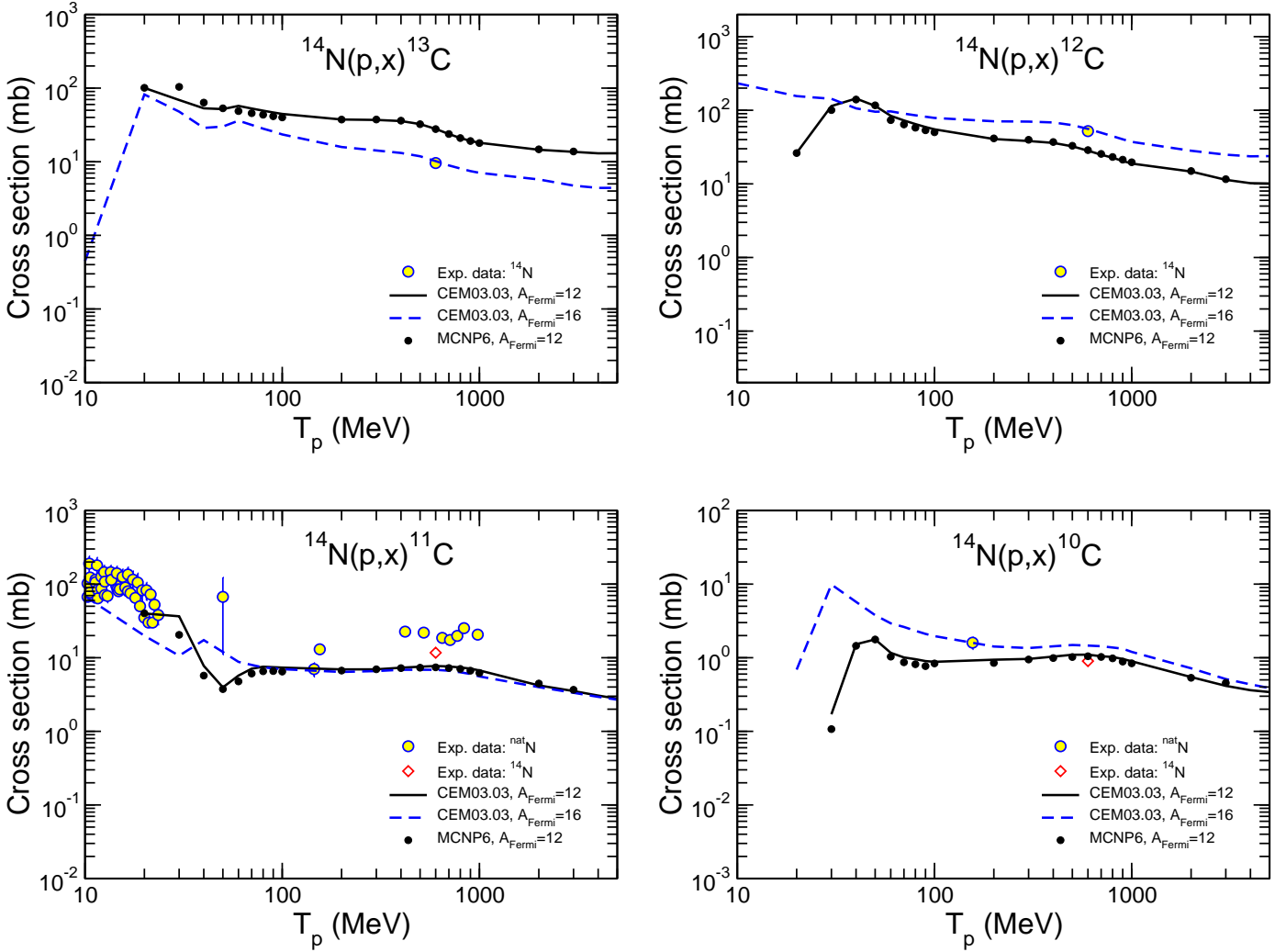}
\caption{Excitation functions for the production 
of  $^{13}$C, $^{12}$C, $^{11}$C, and $^{10}$C from 
p + $^{14}$N calculated with CEM03.03 using the ``standard'' version of the Fermi 
breakup model ($A_{Fermi} = 12$) and with a cut-off value of 
16 for $A_{Fermi}$, as well as with MCNP6 using CEM03.03 
($A_{Fermi} = 12$) compared with experimental data, as indicated. 
Experimental data are from the T16 Lib compilation 
\cite{42}.
%[42].
}
\label{fig:3}
\end{figure*}

{\noindent
 to
what we find
for N and O.
}
 In many cases, we get a better description of the heavy 
fragments when we use $A_{Fermi} = 16$ or 14, and usually we predict a little 
better the light fragments using $A_{Fermi} = 12$. For comparison, for 
Al and Si, we show also excitation functions for the production of all 
complex particles from d to $^4$He, as well as of secondary protons, as we 
found experimental data available for them. Because the absolute values 
of the yields of light fragment production is much lower compared
to the yields of complex particles, and especially of protons, 
the production cross sections of d, t, $^3$He, $^4$He, and especially 
of
p calculated with different values of $A_{Fermi}$ are 
very close to each other. 
This is true also for the production of neutrons; 
although
we do 
not have experimental data for neutron production 
for these reactions.
Generally, emission of nucleons and complex particles are the most determinative in
the calculation of spallation products (heavier residuals) from reactions on medium-mass
nuclei, while LF yields are generally low, and their calculation does not affect
significantly the final cross sections for these heavier products.

Figs. 16 and 17 show 
mass-number dependences of the yield of H, He, Li, Be, B, C, N, and O
isotopes produced in 600 MeV p + $^{16}$O, with a comparison of our CEM results calculated
with $A_{Fermi} = 12$ and 16 with measured data from Ref. 
\cite{47}.
%[47].
There is a relatively good
agreement of both values of $A_{Fermi}$, which does not allow us to choose a preferred value.
The yields of 
$^{11}$B, $^{12}$B, and $^{14}$O with $A_{Fermi} = 16$ agree 
better with the data, while that of
$^{12}$N is predicted better using $A_{Fermi} = 12$.

Fig. 18 shows an example of one more type of nuclear reaction 
characteristic: Atomic-number dependence of the fragment-production cross 
sections from the interactions of $^{20}$Ne (600 MeV/nucleon) with H. 
For this reaction, besides experimental data from Ref. 
\cite{48},
%[48], 
we 
compare to
results calculated with CEM03.03 used as a stand-alone code with 

\begin{figure*}[htb!]
\centering
\includegraphics[width=1.0\textwidth]{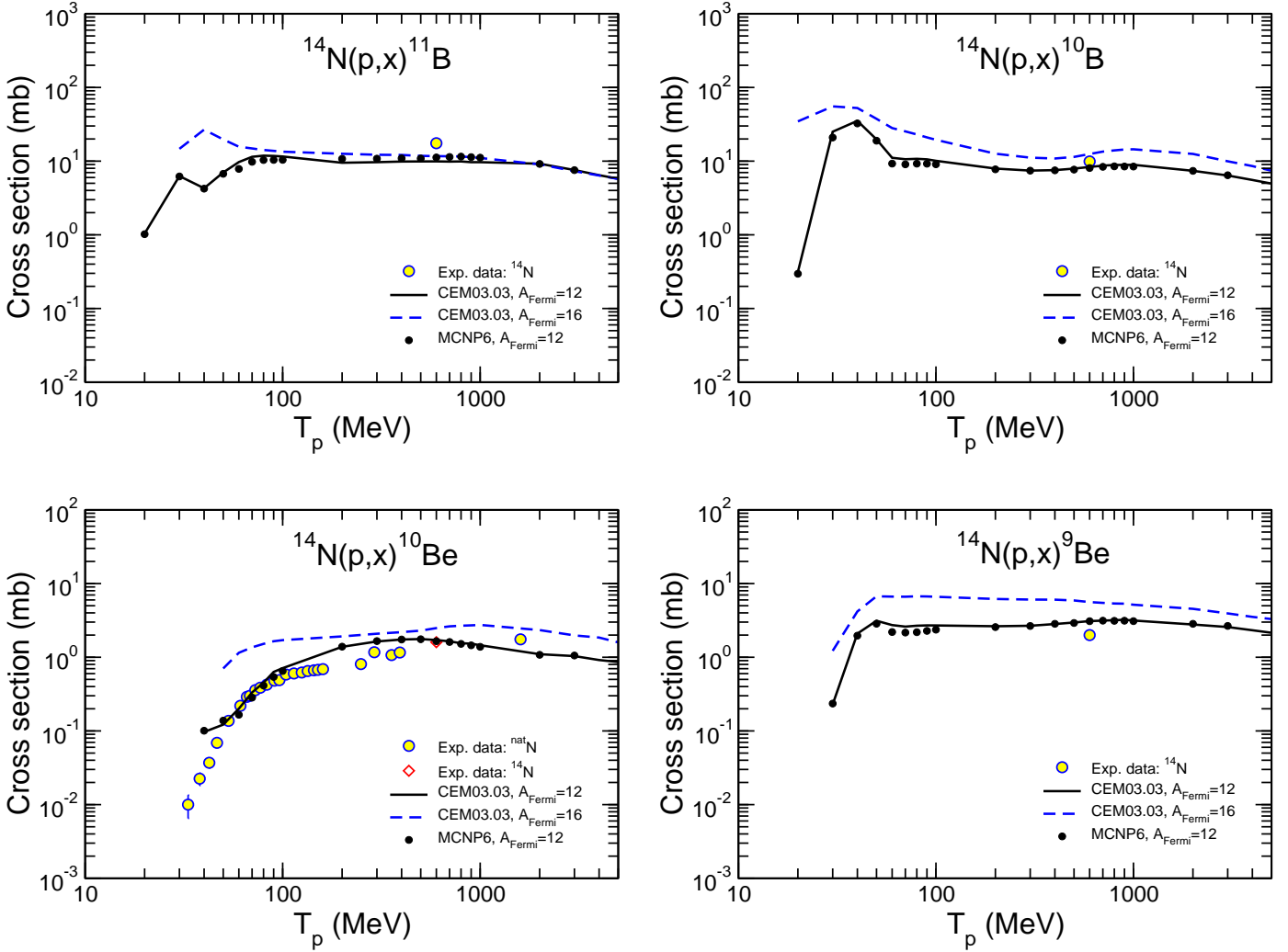}
\caption{The same as in Fig. 3, but for the production 
of  $^{11}$B, $^{10}$B, $^{10}$Be, and $^{9}$Be.}
\label{fig:4}
\end{figure*}

{\noindent 
$A_{Fermi} = 12$ and 16, results by MCNP6 using the CEM03.03 event generator 
with $A_{Fermi} = 12$, as well as results by the NASA semi-empirical nuclear 
fragmentation code NUCFRG2 
\cite{49},
%[49], 
and by a parameterization by Nilsen et al. 
\cite{50}
%[50] 
taken from Tab. III of Ref. 
\cite{48}.
%[48]. 
}
We see that all models agree quite well
with the measured data, especially for LF with $Z > 4$. For LF with 
$Z > 4$, it is difficult to determine from which version of CEM03.03 results 
agree better with the data: the one using $A_{Fermi} = 12$ or the one with 
$A_{Fermi} = 16$.
Light fragments with $Z = 3$ and 4 are described a little better with the 
$A_{Fermi} = 12$ 
version.
As we discuss at the end of the next Section, preequilibrium 
emission described with an extended version of the MEM (not accounted for
in our calculations shown in Fig. 18), can be important and may 
change the final CEM results for this reaction; therefore we 
are not ready
 to make a final decision about which version of 
the Fermi breakup model works better for 
this system.

All the examples  in Figs. 2 to 18 are for reactions 
induced by protons, which at such relatively low incident energies 
are simulated by default in MCNP6 with CEM03.03. 
Figs. 19 and 20 show examples of nucleus-nucleus reactions with light nuclei, 
{\it i.e.}, involving the Fermi breakup model, but simulated in MCNP6 
with LAQGSM03.03. The figures compare experimental 
\cite{51, 48}
%[48, 51] 
$Z$-dependences of products from interactions of 290 MeV/nucleon 
$^{14}$Ne and $^{16}$O with C and Al; 600 MeV/nucleon $^{20}$Ne with C 
and Al; and 400 MeV/nucleon $^{24}$Mg with C and Al with LAQGSM03.03 
results using $A_{Fermi} = 12$ and 16, as well as with results of 
calculations using models of Refs. 
\cite{49, 50, 52},
%[49, 50, 52], 
in the case of 600 MeV/nucleon $^{20}$Ne + C and Al.

The cross sections shown in Figs. 19 and 20 are 
only for the fragmentation of the projectile-nuclei $^{14}$N, $^{16}$O, 
$^{20}$Ne, and $^{24}$Mg; they do not contain contributions from the 
fragmentation of the C and Al target-nuclei. 
For all calculations using all models 
general agreement to the experimental data is quite good.
On the whole, for these particular reactions, the products 
with $Z = 3$ and 4 are described a little better with the 
$A_{Fermi} = 12$ version of LAQGSM, while heavier fragments are 
often  predicted better with $A_{Fermi} = 16$.

\begin{figure*}[htb!]
\centering
\includegraphics[width=1.0\textwidth]{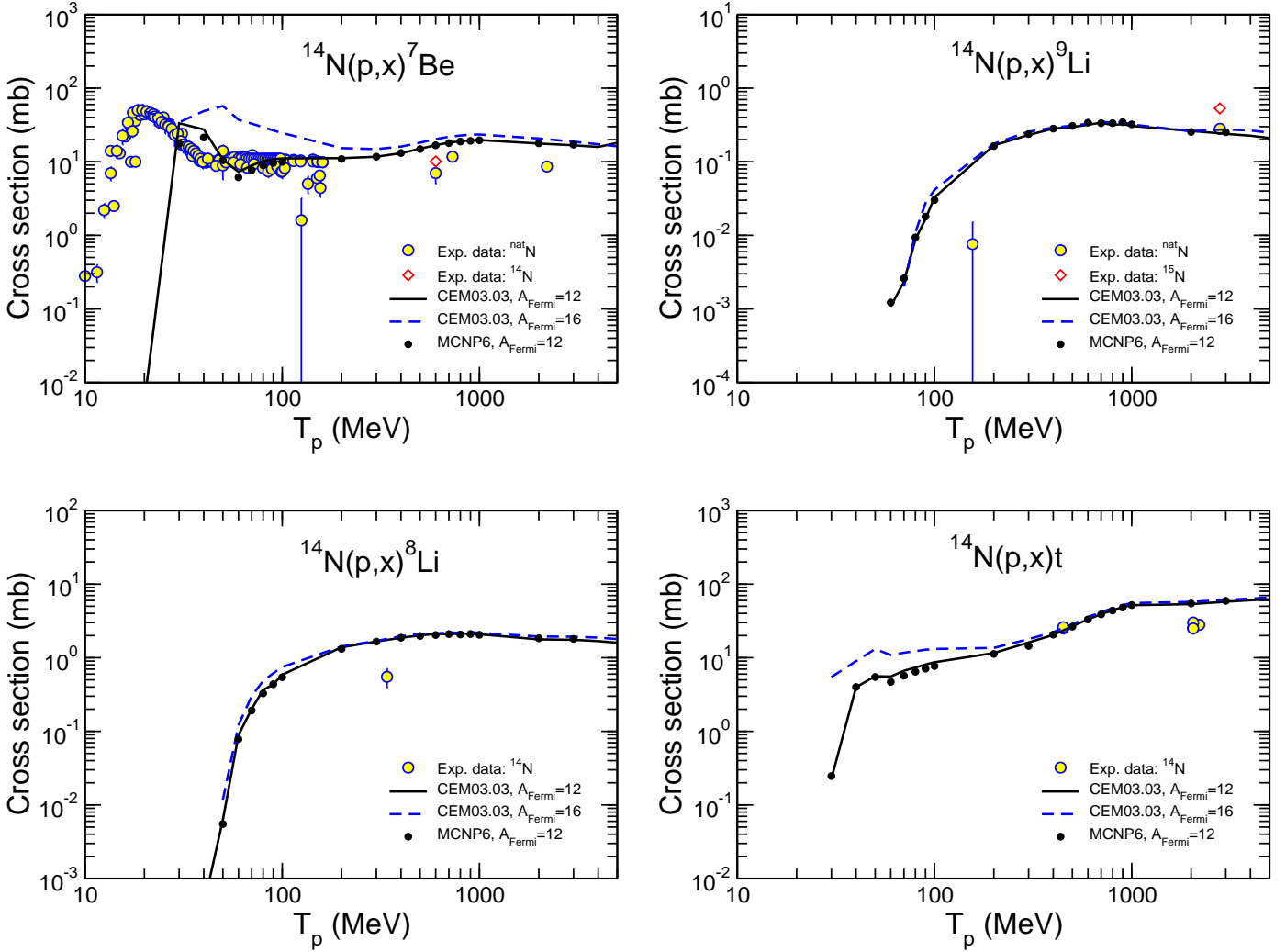}
\caption{The same as in Fig. 3, but for the production 
of  $^{7}$Be, $^{9}$Li, $^{8}$Li, and t.}
\label{fig:5}
\end{figure*}

\subsection{Fragment spectra}
\label{sec-3.2}

This Section presents several examples of particle and LF spectra from 
various proton- and nucleus-induced reactions, chosen so that although 
all of them are fragmentation of light nuclei at intermediate energies, 
they address different reaction mechanisms of fragment production, 
sometimes involving several mechanisms in the production of the same 
LF in a given reaction.

Figs. 21 to 23 show examples of measured particle and LF 
double-differential spectra from p + $^9$Be at 190 and 300 
MeV~\cite{53},
%[53], 
as well as at 392 MeV~\cite{54}
%[54] 
(symbols) compared with our CEM results (histograms).
Because $^9$Be has a mass number $A < A_{Fermi} = 12$, all the LF from 
these reactions are calculated by CEM either as fragments from the Fermi 
breakup of the excited nuclei remaining after the initial INC stage of 
reactions, or as ``residual nuclei'' after emission during INC of several 
particles from the $^9$Be target nucleus. No preequilibrium or/and 
evaporation mechanisms are considered for these reactions by CEM. 
There is
quite a good agreement of the CEM predictions with the measured 
spectra from $^9$Be for all products shown in this example: protons 
(300 MeV p + Be), complex particles (t from 300 MeV p + Be and $^3$He and 
$^4$He from 190 and 392 MeV p + Be), and heavier $^6$He to $^7$Be.

Fig. 24 shows examples of similar LF spectra from a carbon nucleus, 
where only INC and Fermi breakup reaction mechanisms are considered by 
our CEM.
 CEM produces He and Li from these reaction via Fermi breakup after INC, 
while Be and B are probably produced as residual nuclei after emitting 
several nucleons during INC from the carbon target nucleus. The general 
agreement of the CEM predictions with these measured LF spectra is quite 
good, taking into account that no fitting or changing of any parameters 
in CEM was done; we used 
 the fixed version of 
CEM03.03 as implemented in MCNP6.

Fig. 25 shows similar examples of LF spectra,
namely, double-differential spectra at 45 degrees of Li, Be, B, 
and C 
from $^{14}$N and 
$^{16}$O nuclei bombarded with 70 MeV protons.
With these higher mass numbers, 
we performed calculations with CEM03.03 using also 
$A_{Fermi} = 14$ and 16, to see how different values 
 affect the final LF spectra. The general 
agreement of our CEM results with these LF spectra is reasonably good, 
but not quite as good as seen in
Figs. 21 to 24. 
On the whole, for these particular reactions, 
CEM03.03 provides a better agreement with the measured LF spectra 
with
 $A_{Fermi} = 12$.

The examples of LF spectra shown in Figs. 21--25 address 
fragmentation of 
light targets with proton beams.
We present

\begin{figure*}[htb!]
\centering
\includegraphics[width=1.0\textwidth]{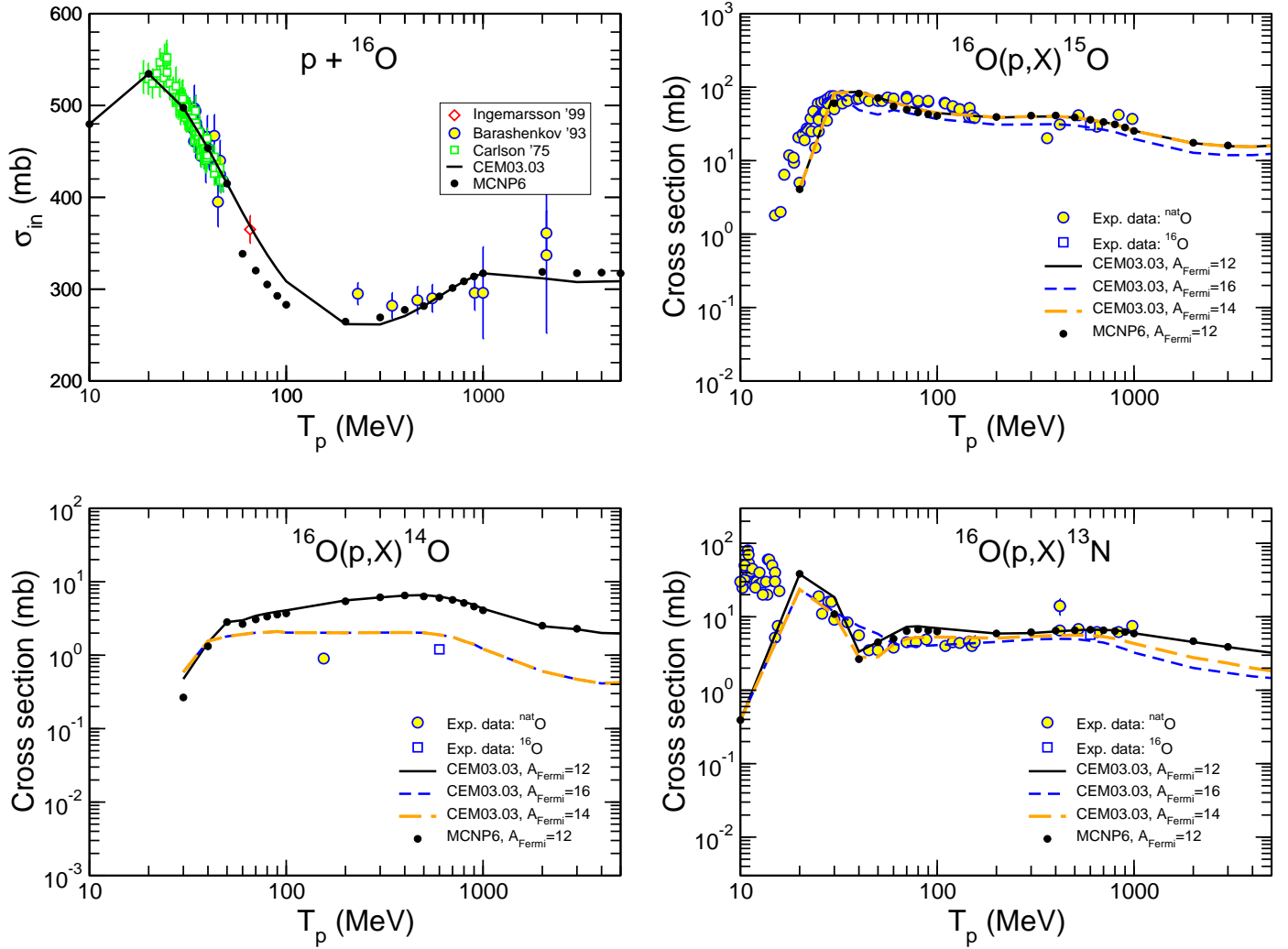}
\caption{Total inelastic cross section and excitation functions for the 
production of $^{15}$O, $^{14}$O, and $^{13}$N from p + $^{16}$O 
calculated with CEM03.03 using the ``standard'' version of the Fermi 
breakup model ($A_{Fermi} = 12$) and with cut-off 
values for $A_{Fermi}$ of 16 and 14, as well as with MCNP6 using CEM03.03 
($A_{Fermi} = 12$) compared with experimental 
data, as indicated. 
Experimental data for inelastic cross sections are from Refs. 
\cite{39, 43, 44},
%[39, 43, 44], 
while the data for excitation functions are from the T16 Lib 
compilation \cite{42}.
%[42].
}
\label{fig:6}
\end{figure*}

{\noindent
 also several examples of LF spectra from 
nucleus-nucleus reactions. Actually, we have already tested MCNP6 
against almost all available 
particle and LF data up to 
$^4$He spectra from various nucleus-nucleus reactions at intermediate 
energies. As a rule, MCNP6 using its LAQGSM03.03 event generator 
describe such spectra quite well (see e.g., Refs. 
\cite{11, 13, 14}
%[11, 13, 14] 
and references therein). The cases 
where there is
 a good agreement with 
experimental data are valuable for MCNP6, to 
verify
 its predictive 
power, but are not so interesting 
for this study,
as they do not address unsolved 
problems in the LAQGSM event-generator. One of the worst
discrepancies
of the LAQGSM LF spectra with available data 
found
from this V\&V of MCNP6 is shown 
below 
in Fig. 26 with dotted lines, namely calculated with the standard 
version of LAQGSM invariant spectra of p, d, t, and $^3$He from 800 
MeV/nucleon $^{20}$Ne + $^{20}$Ne compared with
experimental data from Refs. 
\cite{57, 58}.
%[57, 58]. 
}
$^{20}$Ne nuclei are light enough for the subject of our current work, 
but their mass number $A > A_{Fermi} = 12$, therefore LF can be produced 
by LAQGSM not only with the Fermi breakup model, when the residual 
excited nucleus after INC has a mass number $A < 13$, but also via 
preequilibrium emission and evaporation, as well as final residual 
nuclei after all stages of reactions (see. Fig. 1).

Only LF of high and very high energies were measured in those experiments 
at Bevatron/Bevalac at the Lawrence Berkeley Laboratory 
\cite{57, 58},
%[57, 58], 
and only products from the fragmentation of the bombarding nuclei were 
detected. LAQGSM can reproduce such high-energy portions of spectra only 
with its coalescence model, as the Fermi breakup model would provide LF of 
lower energies, while the preequilibrium emission and 

\begin{figure*}[htb!]
\centering
\includegraphics[width=1.0\textwidth]{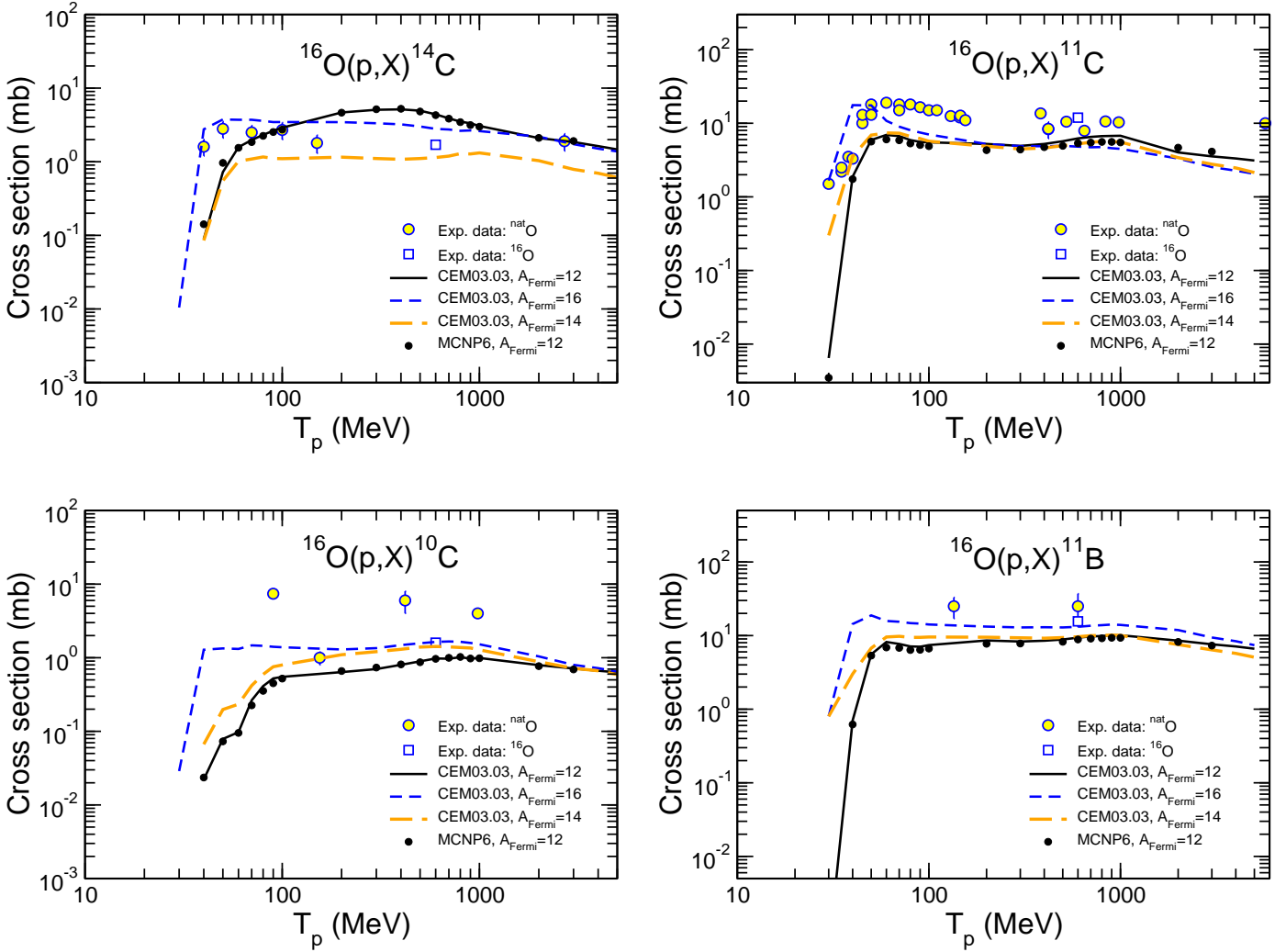}
\caption{Excitation functions for the 
production of $^{14}$C, $^{11}$C, $^{10}$C, and $^{11}$B from p + $^{16}$O 
calculated with CEM03.03 using the ``standard'' version of the Fermi 
breakup model ($A_{Fermi} = 12$) and with cut-off 
values for $A_{Fermi}$ of 16 and 14, as well as with MCNP6 using CEM03.03 
($A_{Fermi} = 12$) compared with experimental 
data, as indicated. 
Experimental data are from the T16 Lib 
compilation \cite{42}.
%[42].
}
\label{fig:7}
\end{figure*}

{\noindent
evaporation would 
provide much lower LF energies; the energies of the LF produced as final
``residual nuclei'' after all other stages of reactions would be even much 
lower. 
}
In other words, the experimental data from Refs. 
\cite{57, 58}
%[57, 58] 
are very convenient to 
test the coalescence model in LAQGSM.

As noted in Section 2.2, LAQGSM uses fixed values for $p_c$ as 
determined by Eq. (1).
Results obtained 
with such ``standard'' values for $p_c$ are shown in Fig. 26 with dotted 
lines: We see that these LAQGSM spectra underestimate the measured data, 
suggesting that we need to use higher values for $p_c$, 
at least for this particular reaction. As a second test of 
$p_c$ values, we try to use for this reaction the values
\begin{eqnarray}
p_c(d) & = & 115 \mbox{ MeV/c ;} \nonumber \\
p_c(t) & = & p_c(^3{\mbox He}) = \\
= p_c(^4{\mbox He}) & = & 175 \mbox{ MeV/c ,} \nonumber 
\label{e3}
\end{eqnarray}
found to work the best in CEM03.03 in the 300 MeV $\leq T < 1$ GeV region of 
incident energies.
Results obtained with these values for $p_c$ are shown in Fig. 26 with 
dashed lines:
We see that values of $p_c$ defined by Eq. (3) provide too many 
high energy LF, {\it i.e.}, these values are too big to provide the best 
results for LF spectra calculated by LAQGSM for this particular 
reaction. Finally, we try some intermediate $p_c$ values:
\begin{eqnarray}
p_c(d) & = & 120 \mbox{ MeV/c ;} \nonumber \\
p_c(t) & = & p_c(^3{\mbox He}) = \\
= p_c(^4{\mbox He}) & = & 140 \mbox{ MeV/c .} \nonumber 
\label{e1}
\end{eqnarray}
Results calculated with these values are shown with solid 
lines in Fig. 26: They agree much better with the measured spectra of d, t, 
and $^3$He from this reaction than the previous two sets of results. 
Note that the aim of our work is not to fine-tune the parameters used 
by the coalescence model in our CEM and LAQGSM event generators. We may 
consider such a fine-tuning at a later stage, after we complete our work 
on extension of the preequilibrium model to account for possible emission 
of LF heavier than
$^4$He at the preequilibrium stage of reactions,
 discussed below. 
Here, we just show that although the standard versions of our CEM 
and LAQGSM 
event generators for MCNP6 provide an overall good agreement 
of calculated spectra and yields of products from various reactions, a 
fine-tuning of some of their parameters would allow improving further the 
agreement of calculated results with available experimental data.

Finally, we
mention briefly our preliminary results 
from 
recent work
\cite{59, 60}
%[59, 60] 
to extend the CEM and LAQGSM for accounting possible emission of LF heavier 
than $^4$He (up to $^{28}$Mg)

\begin{figure*}[htb!]
\centering
\includegraphics[width=1.0\textwidth]{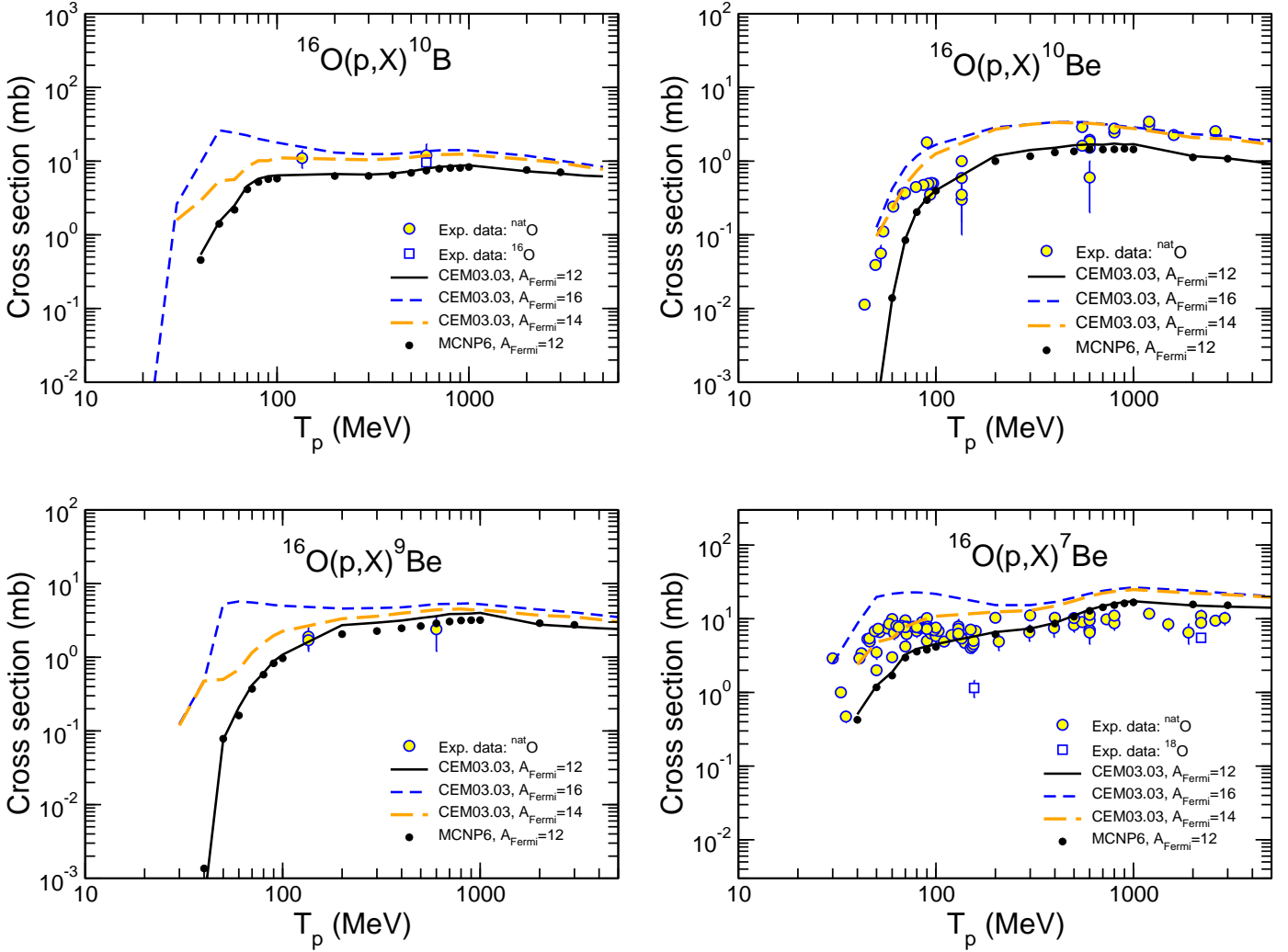}
\caption{The same as in Fig. 7, but 
for the 
production of $^{10}$B, $^{10}$Be, $^{9}$Be, and $^{7}$Be.
}
\label{fig:8}
\end{figure*}

{\noindent
at the preequilibrium stage of nuclear 
reactions. 
}
Fig. 27 shows an example of such results, namely, $^6$Li spectra 
from the reaction 200 MeV p + $^{27}$Al measured by Machner et al. 
\cite{61}
%[61] 
(symbols) compared with our preliminary results by an extended version of 
CEM, as described in Refs. 
\cite{59, 60}
%[59, 60] 
(solid lines) and with results by the standard unmodified CEM
(dashed lines). 
The
aluminum target is relatively light, {\it i.e.}, such 
reactions are completely in the scope of this study.
However, 
in comparison with lighter targets like C, N, and O discussed above, Al 
is heavier, the mass number of excited nuclei produced after INC from such 
reactions 
is mostly
 higher than 12, therefore CEM uses 
the preequilibrium and evaporation models to calculate this reaction, 
in addition to INC, coalescence, and Fermi breakup models (see Fig. 1).

The $^6$Li spectra calculated with the standard version of CEM as 
implemented at present in MCNP6 shown with dashed histograms came 
mostly from evaporation of $^6$Li from compound nuclei, and also contain 
a small contribution from Fermi breakup of excited nuclei with 
$A \leq A_{Fermi} = 12$ produced in a few cases after INC from Al-target, 
at such a relatively low incident energy of only 200 MeV. As expected, 
evaporation of $^6$Li from compound nuclei together with a small 
contribution from Fermi breakup of nuclei with $A \leq A_{Fermi} = 12$ 
produced after INC in this reaction do not provide enough high-energy 
LF emission, and the calculated $^6$Li spectra do not extend to high 
energies and are below the measured data. Extension of the preequilibrium 
model used by CEM and LAQGSM to account for emission of LF heavier than 
$^4$He as described in Refs. 
\cite{59, 60},
%[59, 60], 
allows us to produce energetic LF from such reactions, improving the 
agreement with many measured LF spectra we 
tested so far. Of course, 
preequilibrium emission of LF is most important for medium and heavy 
target nuclei.
But as we see in this example,  it  also
affects significantly such relatively light nuclei 
as Al. Our work on extending the
 preequilibrium model in CEM and 
LAQGSM is incomplete.
 Results from this study
will be published in the future. 
After completing this work,
we may consider a fine-tuning of the $A_{Fermi}$ 
parameter of the Fermi breakup model and of the
coalescence model parameters used by the CEM and LAQGSM event 
generators of MCNP6.

\begin{figure*}[htb!]
\centering
\includegraphics[width=1.0\textwidth]{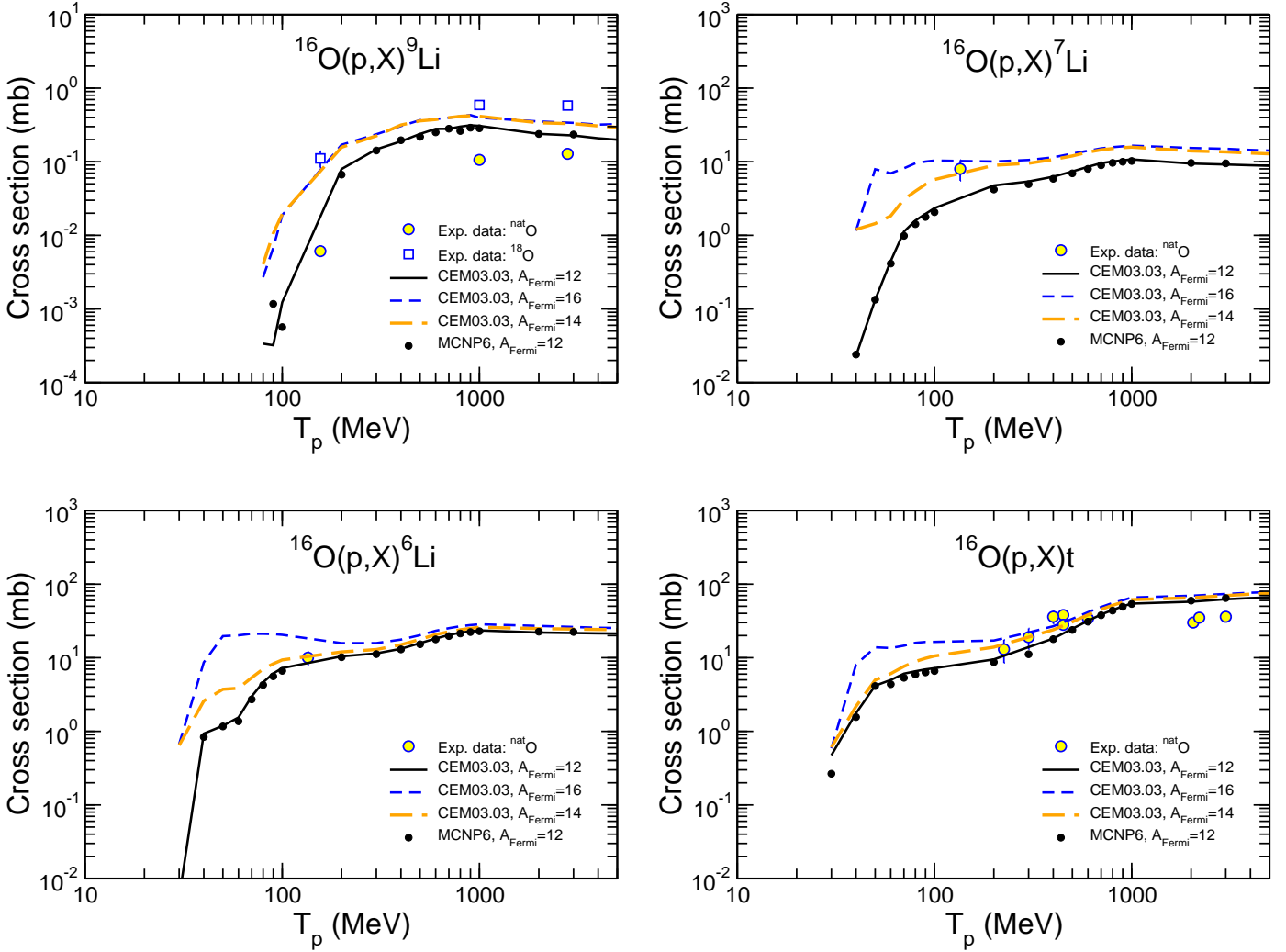}
\caption{The same as in Fig. 7, but 
for the 
production of $^{9}$Li, $^{7}$Li, $^{6}$Li, and t.
}
\label{fig:9}
\end{figure*}

{\noindent
 }

\section{Conclusion}
\label{sec-4}

Various fragmentation reactions induced by protons and light nuclei 
of energies around 1 GeV/nucleon and below on light target nuclei are 
studied with the latest Los Alamos Monte Carlo transport code MCNP6 
and with its cascade-exciton model (CEM) and Los Alamos version of 
the quark-gluon string model (LAQGSM) event generators, version 
03.03, used as stand-alone codes. On the whole, MCNP6 and its CEM 
and LAQGSM event generators describe quite well all the reactions 
we tested here, providing good enough agreement with available 
experimental data. This is especially important for calculations 
of cross sections of arbitrary products as functions of incident 
projectile energies, {\it i.e.}, excitation functions, one of the most 
difficult tasks for any nuclear reaction model. Our current results 
show a good prediction by MCNP6 and CEM03.03, used as a stand-alone 
code, of a large variety of excitation functions for products from 
proton-induced reactions on N, O, Al, and Si. An older version 
of CEM, CEM95, was able to predict reasonably well most excitation 
functions for medium and heavy nuclei-targets, but had big problems in 
calculating some excitation functions for light nuclei 
\cite{62}.
%[62].

CEM and LAQGSM assume that intermediate-energy fragmentation reactions 
on light nuclei occur generally in two stages. The first stage is the 
intranuclear cascade (INC), followed by the second, Fermi breakup 
disintegration of light excited residual nuclei produced after INC. 
Both CEM and LAQGSM  also account for coalescence of light fragments 
(complex particles) up to $^4$He from energetic nucleons emitted during 
INC.

\begin{figure*}[htb!]
\centering
\includegraphics[width=1.0\textwidth]{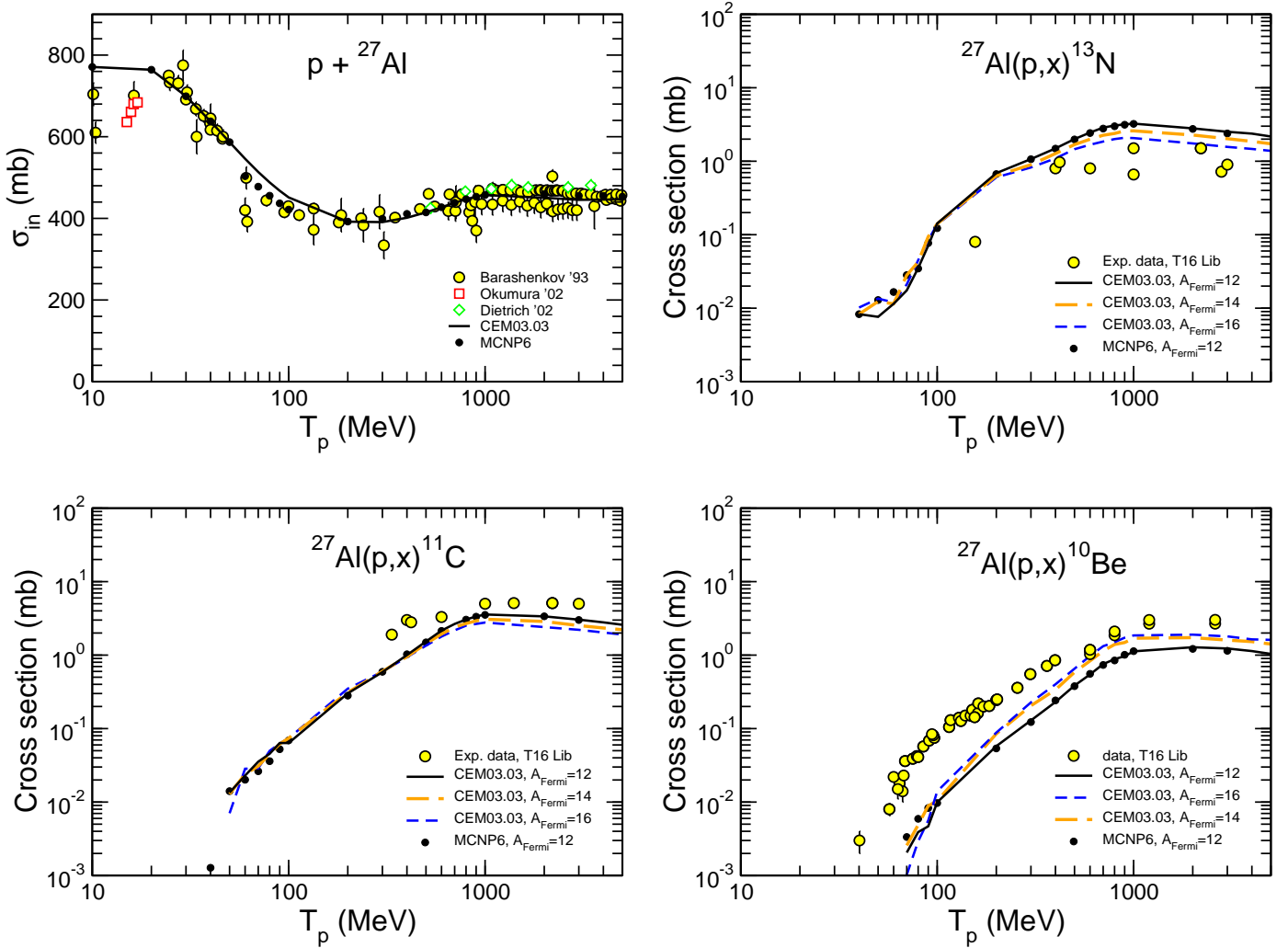}
\caption{Total inelastic cross section and excitation functions for 
the production of $^{13}$N, $^{11}$C, and $^{10}$Be, from 
p + $^{27}$Al calculated with CEM03.03 using the ``standard'' 
version of the Fermi breakup model ($A_{Fermi} = 12$) and with 
cut-off values for $A_{Fermi}$ of 16 
 and 14, as well as with MCNP6 
using CEM03.03 ($A_{Fermi} = 12$) 
compared with experimental data, as indicated. 
Experimental data for inelastic 
cross sections are from Refs. 
\cite{39, 45, 46},
%[39, 45, 46], 
while the data for excitation functions are from the T16 Lib 
compilation 
\cite{42}.
%[42].
}
\label{fig:10}
\end{figure*}

We investigate the validity and performance of MCNP6, CEM, and LAQGSM 
in simulating fragmentation reactions at intermediate energies. We find 
that while the
fixed ``default'' versions of CEM03.03 and LAQGSM03.03 
in MCNP6 
provide
reasonably good predictions for all reactions 
tested here, a fine-tuning of the $A_{Fermi}$ parameter in the Fermi breakup 
model and of 
momentum cut-off
parameters in the coalescence model may provide a better 
description of some experimental data. We may consider such a fine-tuning 
of these and other CEM and LAQGSM parameters later, after we complete our 
work 
\cite{59, 60}
%[59, 60] 
on extending the
 preequilibrium model in CEM and LAQGSM to account 
for possible emission of light fragments heavier than $^4$He (up to 
$^{28}$Mg) at this  stage of reactions.

\vspace*{5mm}
{\noindent
{\bf Acknowledgments}
}

We are grateful to our colleagues, Drs. Konstantin K. Gudima and 
Arnold J. Sierk for a long and very fruitful collaboration with 
us and for several useful discussions of the results presented here.

We thank Drs. Masayaki Hagiwara, Shoji Nagamiya, Toshiya Sanami, 
Nikolai M. Sobolevsky, and Yusuke Uozumi for sending us their 
publications or/and files with numerical values of their experimental 
data we use in our work.

We are grateful to Drs. Lawrence J. Cox and Avneet Sood of LANL and
to Prof. Akira Tokuhiro of University of Idaho for
encouraging discussions and support.

Last but not least, we thank Dr. Roger L. Martz for a careful reading 
of our manuscript and useful suggestions on its improvement.

This study was carried out under the auspices of the National Nuclear 
Security Administration of the U.S. Department of Energy at Los Alamos 
National Laboratory under Contract No. DE-AC52-06NA253996.

This work is supported in part (for L.M.K) by the M. Hildred Blewett 
Fellowship of the American Physical Society, 

%\clearpage

\begin{figure*}[htb!]
\centering
\includegraphics[width=1.0\textwidth]{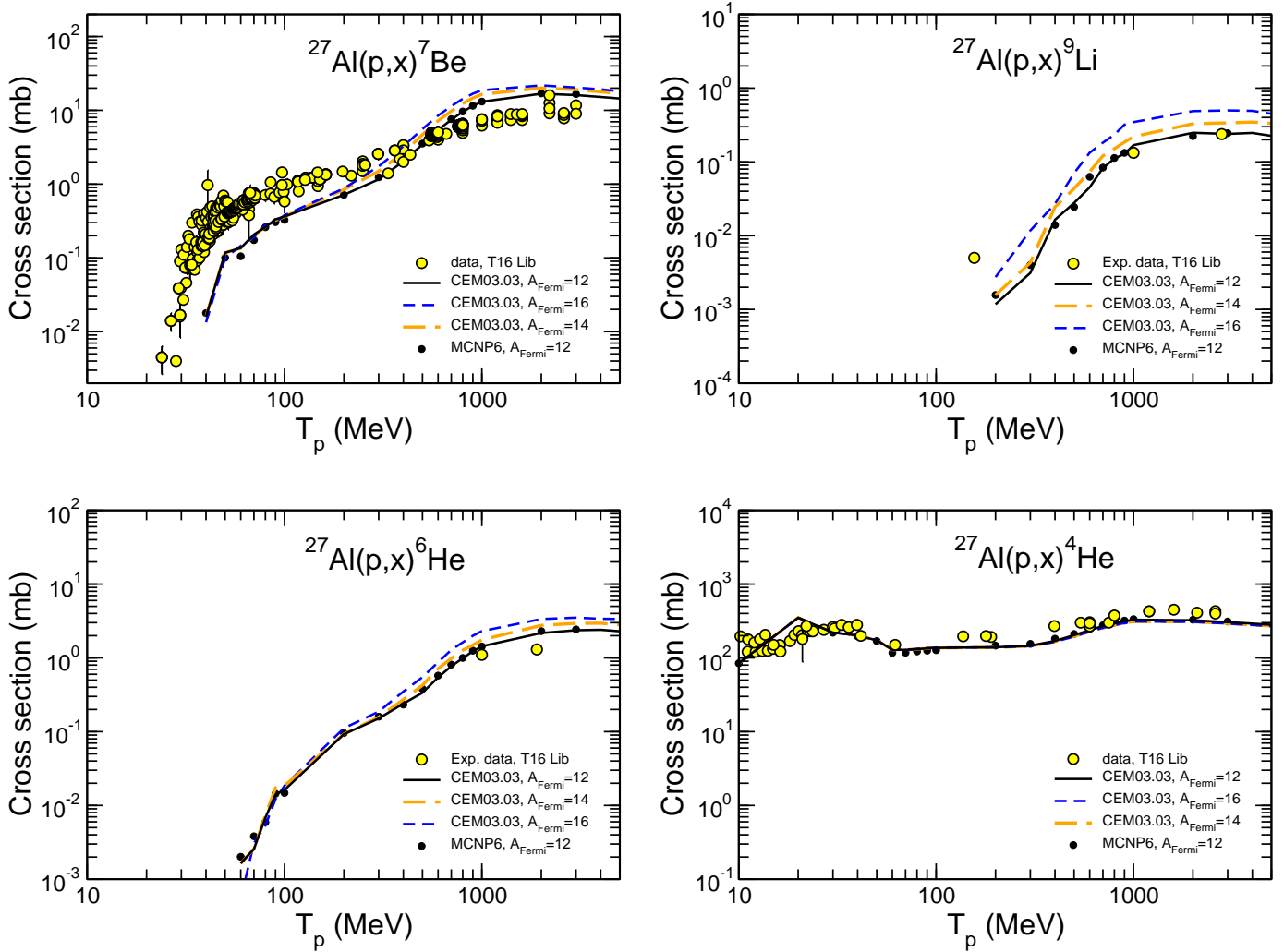}
\caption{Excitation functions for the production of 
$^7$Be, $^9$Li, $^6$He, and $^4$He
from p + $^{27}$Al calculated with CEM03.03 using the ``standard"
version of the Fermi breakup model ($A_{Fermi} = 12$) 
and with cut-off values for $A_{Fermi}$ of 16  and 14, as well as with MCNP6 
using CEM03.03 ($A_{Fermi} = 12$) 
compared with experimental data, as indicated. 
Experimental data are from the T16 Lib 
compilation 
\cite{42}.
%[42].
}
\label{fig:11}
\end{figure*}

{\noindent
www.aps.org.
}

%\vspace*{1.5cm}

\vspace*{5mm}
{\noindent
{\bf References}
}

\clearpage 

\begin{figure*}[htb!]
\centering
\includegraphics[width=1.0\textwidth]{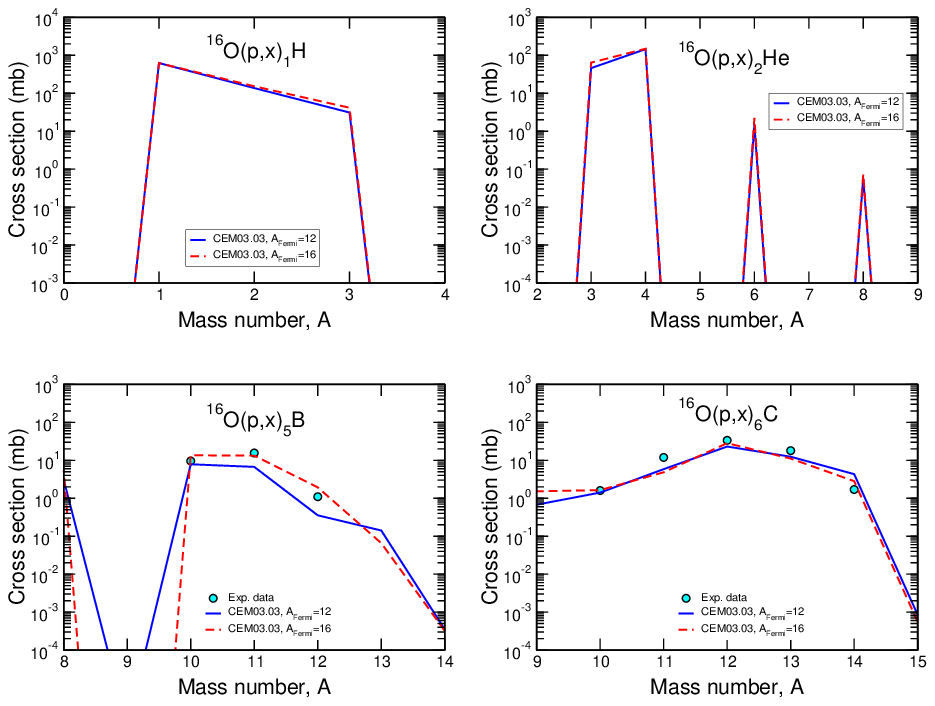}
\vspace*{2mm}
\caption{
Examples of measured particle and LF production cross 
sections from p + $^{16}$O at 600 MeV 
\cite{47}
%[47] 
(symbols) compared with our CEM results for a Fermi breakup cut-off of 
$A \leq 16$ and $A \leq 12$, as indicated. 
All the LF from these reactions are calculated by CEM either as final 
products (residual nuclei) after all possible stages of reaction or via 
Fermi breakup after INC (Fermi breakup is used for nuclei with 
$A < 13$ or $A < 17$ instead of using preequilibrium emission and/or 
evaporation of particles).
}
\label{fig:16}
\end{figure*}

\clearpage 

\begin{figure*}[htb!]
\centering
\includegraphics[width=1.0\textwidth]{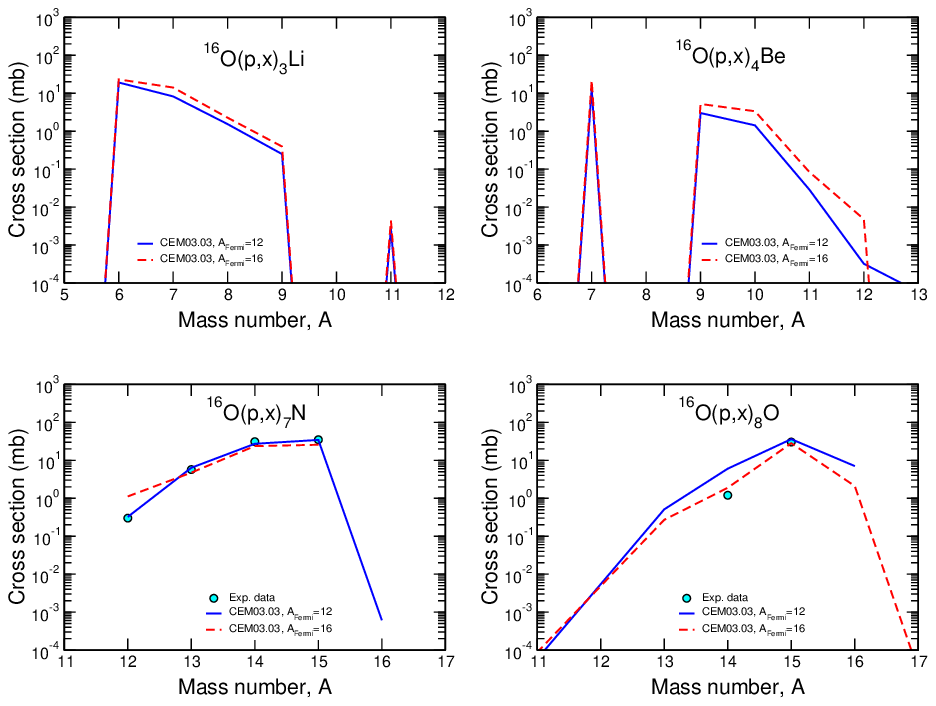}
\vspace*{0.5mm}
\caption{The same as in Fig. 16, but 
for the 
production of Li, Be, N, and O isotopes.
}
\label{fig:17}
\end{figure*}

\clearpage 

\begin{figure*}[t!]
\vspace*{2mm}
\centering
\includegraphics[width=0.75\textwidth]{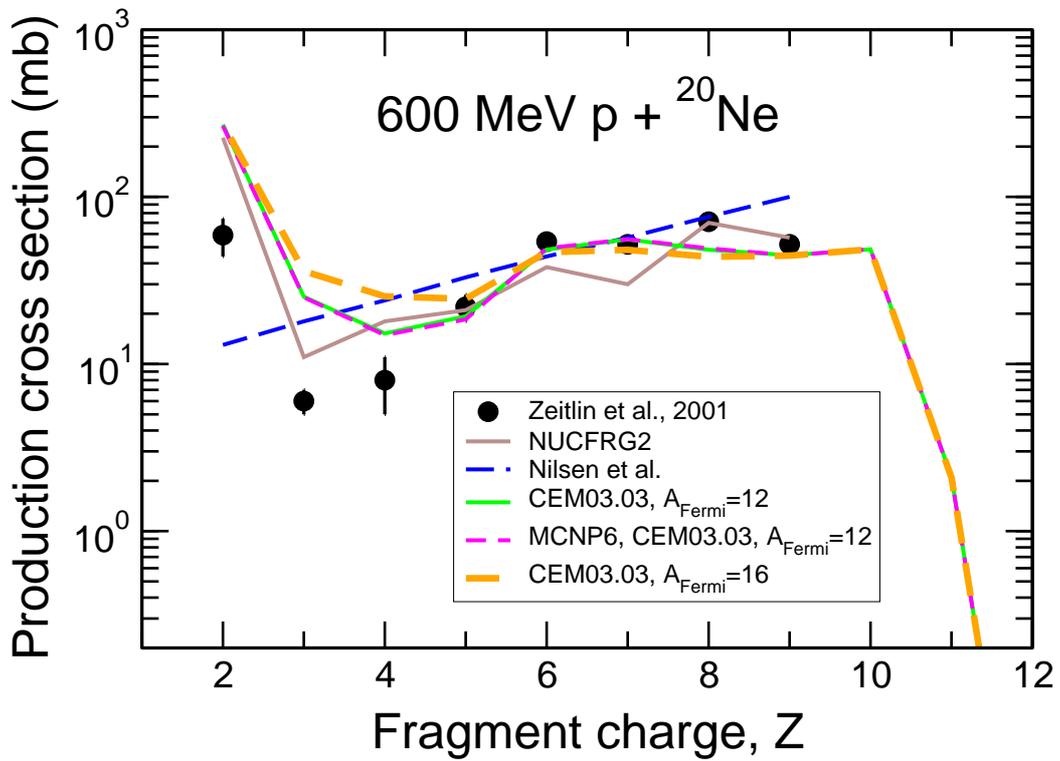}
\vspace{4mm}
\caption{Atomic-number dependence of the fragment-production cross 
sections from the interactions of $^{20}$Ne of 600 MeV/nucleon 
with H. Experimental data (circles) are by Zeitlin et al. 
\cite{48}.
%[48]. 
For comparison, 
results by the NASA 
semi-empirical nuclear fragmentation code NUCFRG2 
\cite{49} and 
%[49], 
from a parameterization 
by Nilsen et al. 
\cite{50}
%[50] 
taken from Tab. III of Ref. 
\cite{48} are shown as well, as indicated.
%[48]. 
Our results by CEM03.03 using the ``standard'' version of the Fermi 
breakup model ($A_{Fermi} = 12$) and
$A_{Fermi} = 16$, as well as
MCNP6  calculations using CEM03.03 ($A_{Fermi} = 12$) are 
plotted with different lines, as indicated.
}
\label{fig:18}
\end{figure*}

\clearpage 

\begin{figure*}[htb!]
\centering
\includegraphics[width=1.0\textwidth]{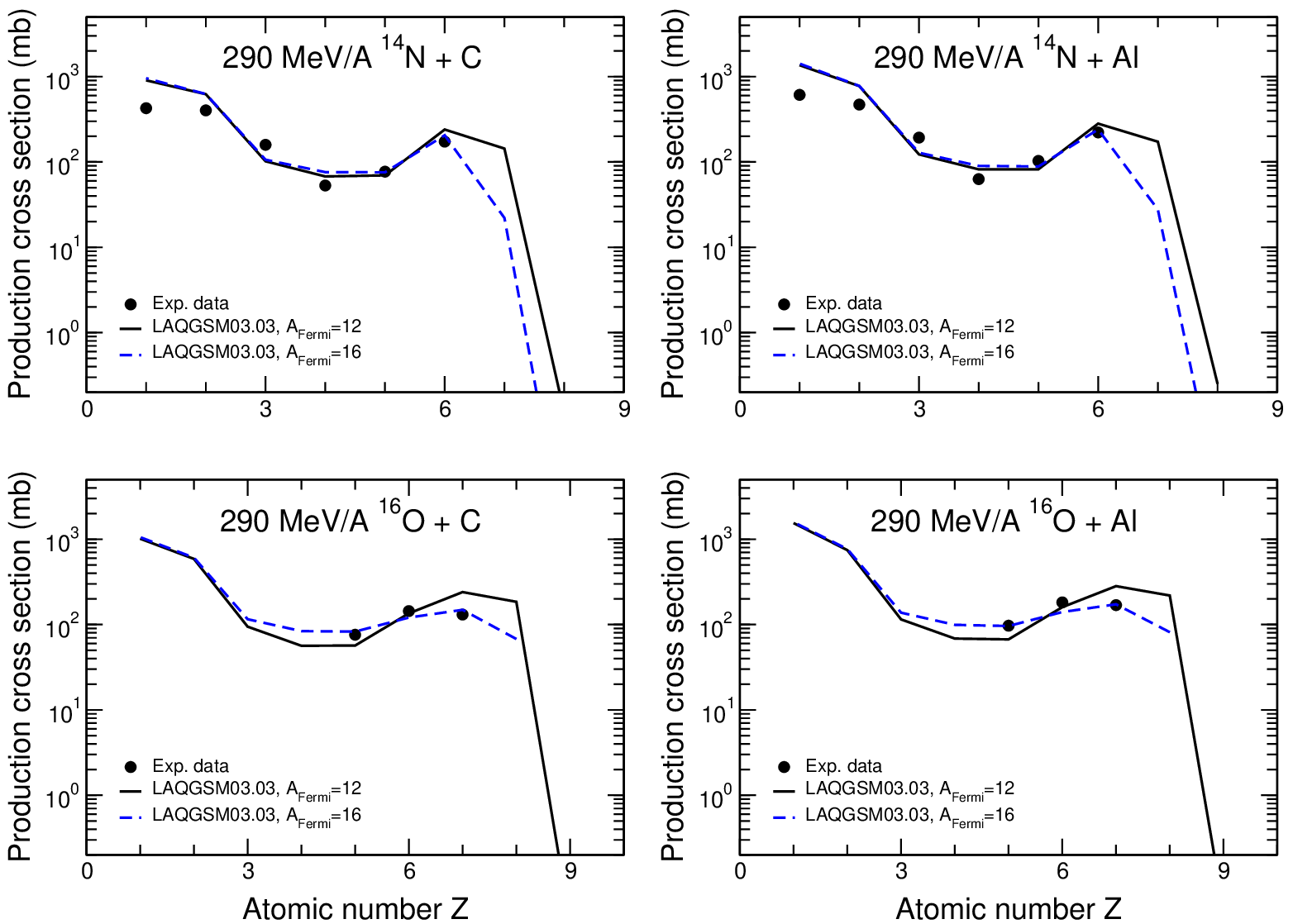}
\vspace*{0.5mm}
\caption{Atomic-number dependence of the fragment-production cross sections 
from the interactions of 290 MeV/nucleon $^{14}$Ne and $^{16}$O with C and Al. 
Experimental data (circles) are by Zeitlin et al. 
\cite{51}.
%[51]. 
Our results by LAQGS03.03 using the ``standard'' version of the Fermi breakup 
model ($A_{Fermi} = 12$) are shown with solid lines, and for a cut-off 
value for $A_{Fermi}$ of 16, with dashed lines, as indicated.
}
\label{fig:19}
\end{figure*}

\clearpage

\begin{figure*}[htb!]
\centering
\includegraphics[width=1.0\textwidth]{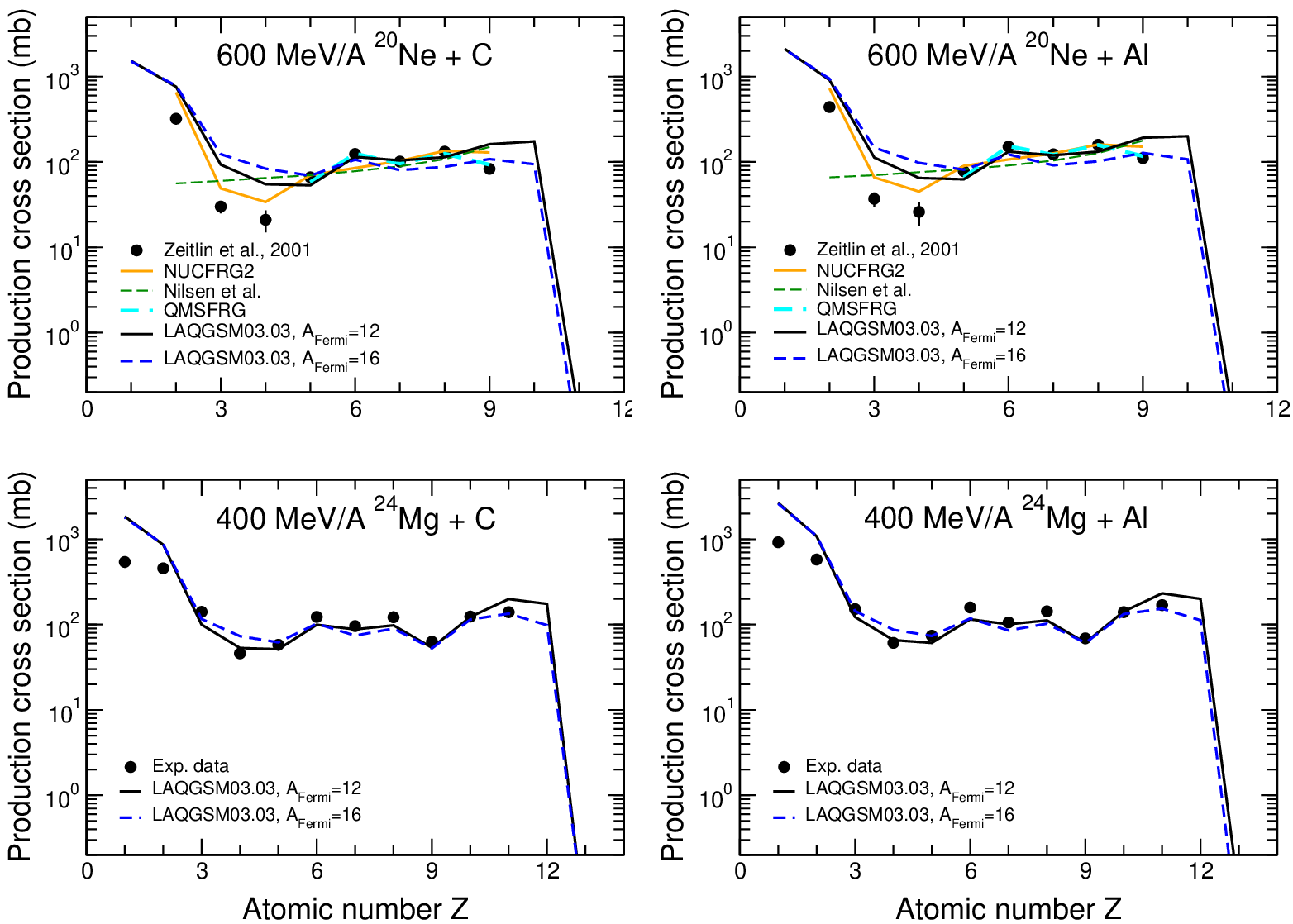}
\vspace*{0.5mm}
\caption{Atomic-number dependence of the fragment-production cross sections 
from the interactions of 
600 MeV/nucleon $^{20}$Ne with C and Al, and 400 MeV/nucleon $^{24}$Mg 
with C and Al.
Experimental data (circles) are by Zeitlin et al. 
\cite{51, 48}.
%[48, 51]. 
For comparison, at 600 MeV/A, results by the NASA semi-empirical nuclear 
fragmentation code NUCFRG2 
\cite{49}
%[49] 
and the microscopic abrasion-ablation model QMSFRG 
\cite{52},
%[52], 
as well as from a parameterization by Nilsen et al. 
\cite{50}
%[50] 
taken from Tabs. III and IV of Ref. 
\cite{48}
%[48] 
are shown with different lines, as indicated. 
Our results by LAQGS03.03 using the ``standard'' version of the Fermi breakup 
model ($A_{Fermi} = 12$) are shown with solid lines, and for a cut-off 
value for $A_{Fermi}$ of 16, with dashed lines, as indicated.
}
\label{fig:20}
\end{figure*}

\clearpage

\begin{figure*}[htb!]
\centering
\includegraphics[width=1.0\textwidth]{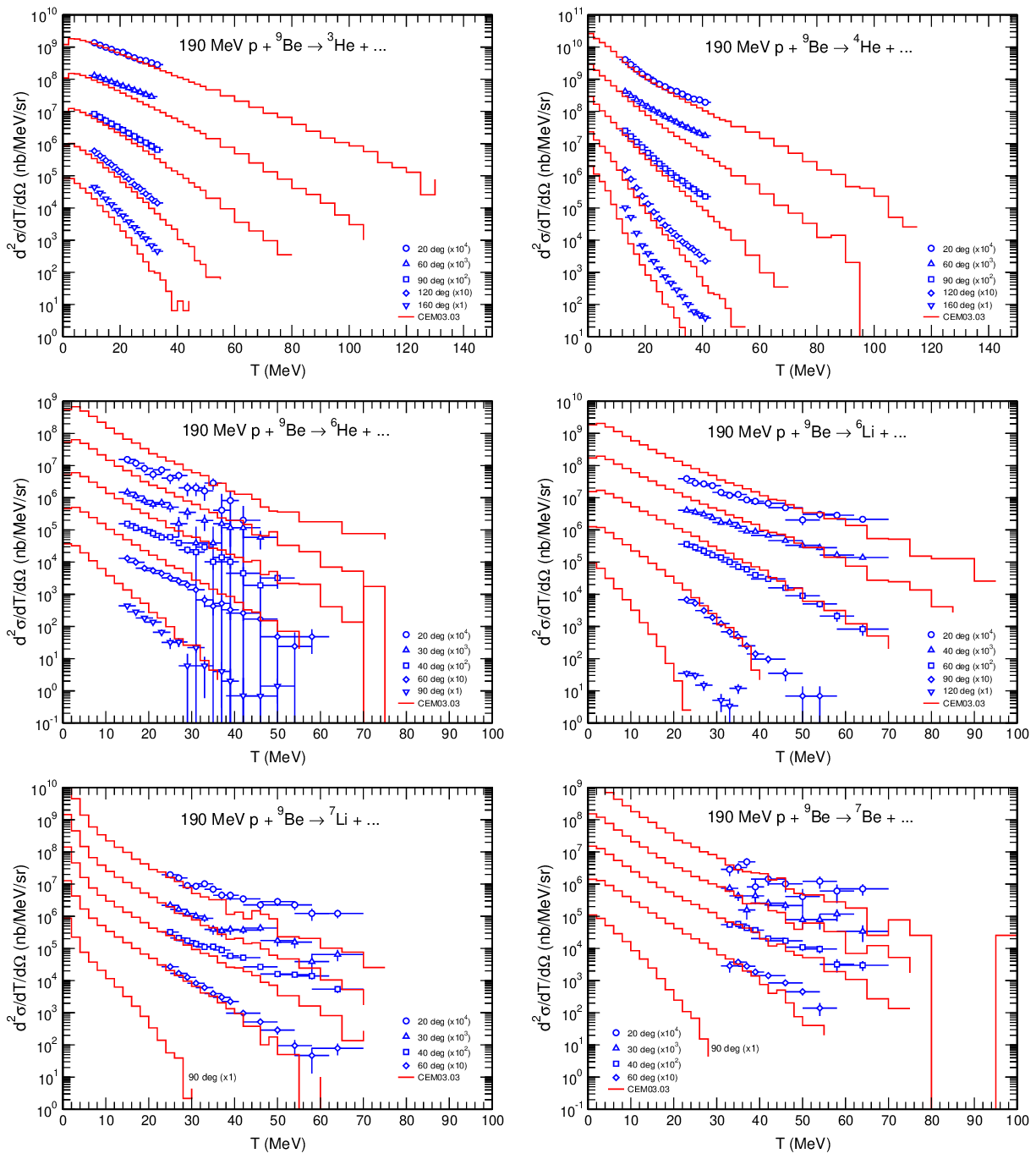}
%\vspace*{2mm}
\caption{
Examples of measured particle and LF double-differential spectra from 
p + $^9$Be at 190 MeV 
\cite{53},
%[53],
compared with our CEM results (histograms).
}
\label{fig:21}
\end{figure*}

\clearpage

\begin{figure*}[htb!]
\centering
\includegraphics[width=1.0\textwidth]{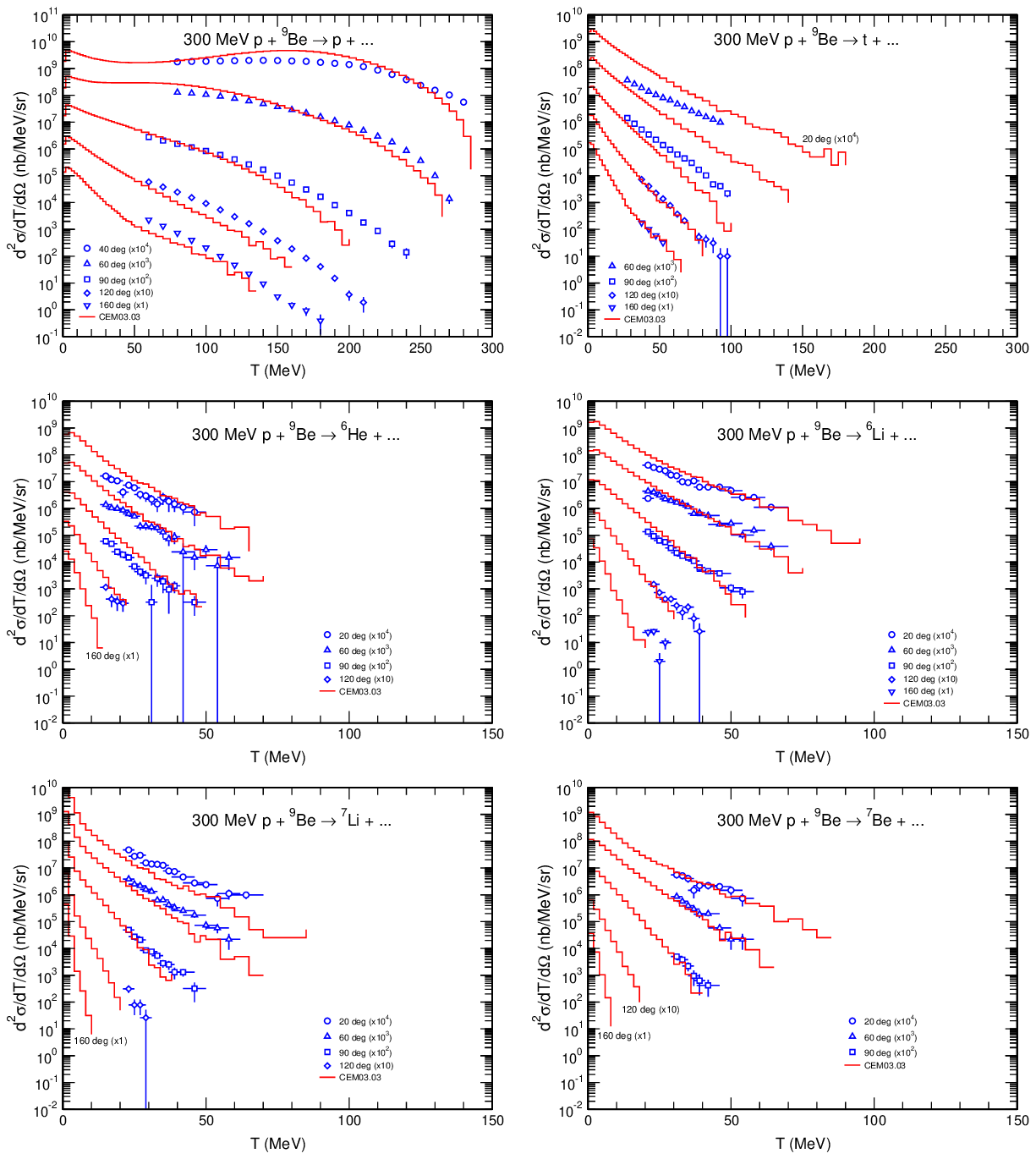}
%\vspace*{2mm}
\caption{
Examples of measured particle and LF double-differential spectra from 
p + $^9$Be at 300 MeV 
\cite{53},
%[53],
compared with our CEM results (histograms).
}
\label{fig:22}
\end{figure*}

\clearpage

\begin{figure*}[htb!]
\centering
\includegraphics[width=1.0\textwidth]{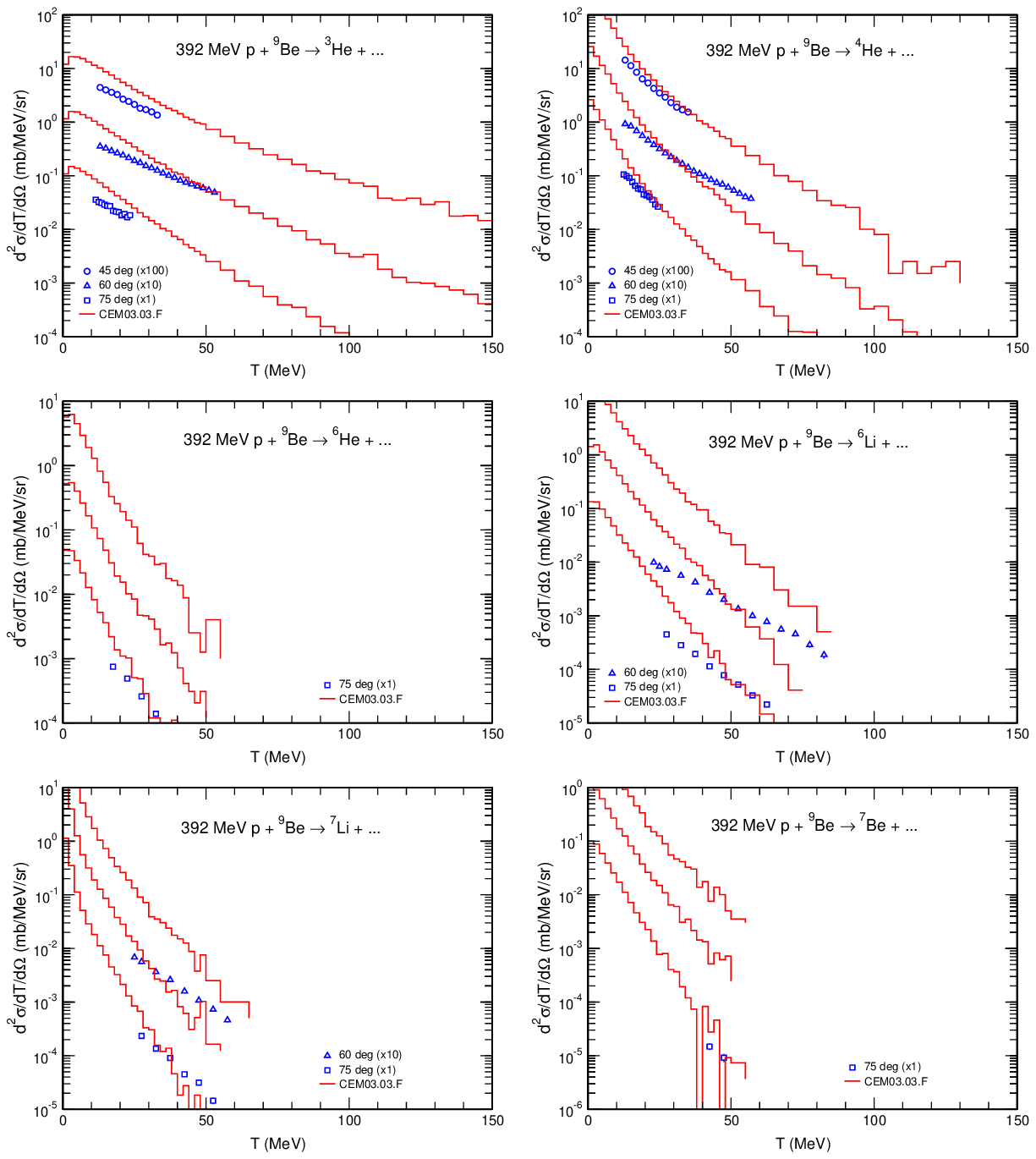}
%\vspace*{2mm}
\caption{
Examples of measured particle and LF double-differential spectra from 
p + $^9$Be at 392 MeV 
\cite{54},
%[54],
compared with our CEM results (histograms).
}
\label{fig:23}
\end{figure*}

\clearpage

\begin{figure*}[htb!]
\centering
\includegraphics[width=1.0\textwidth]{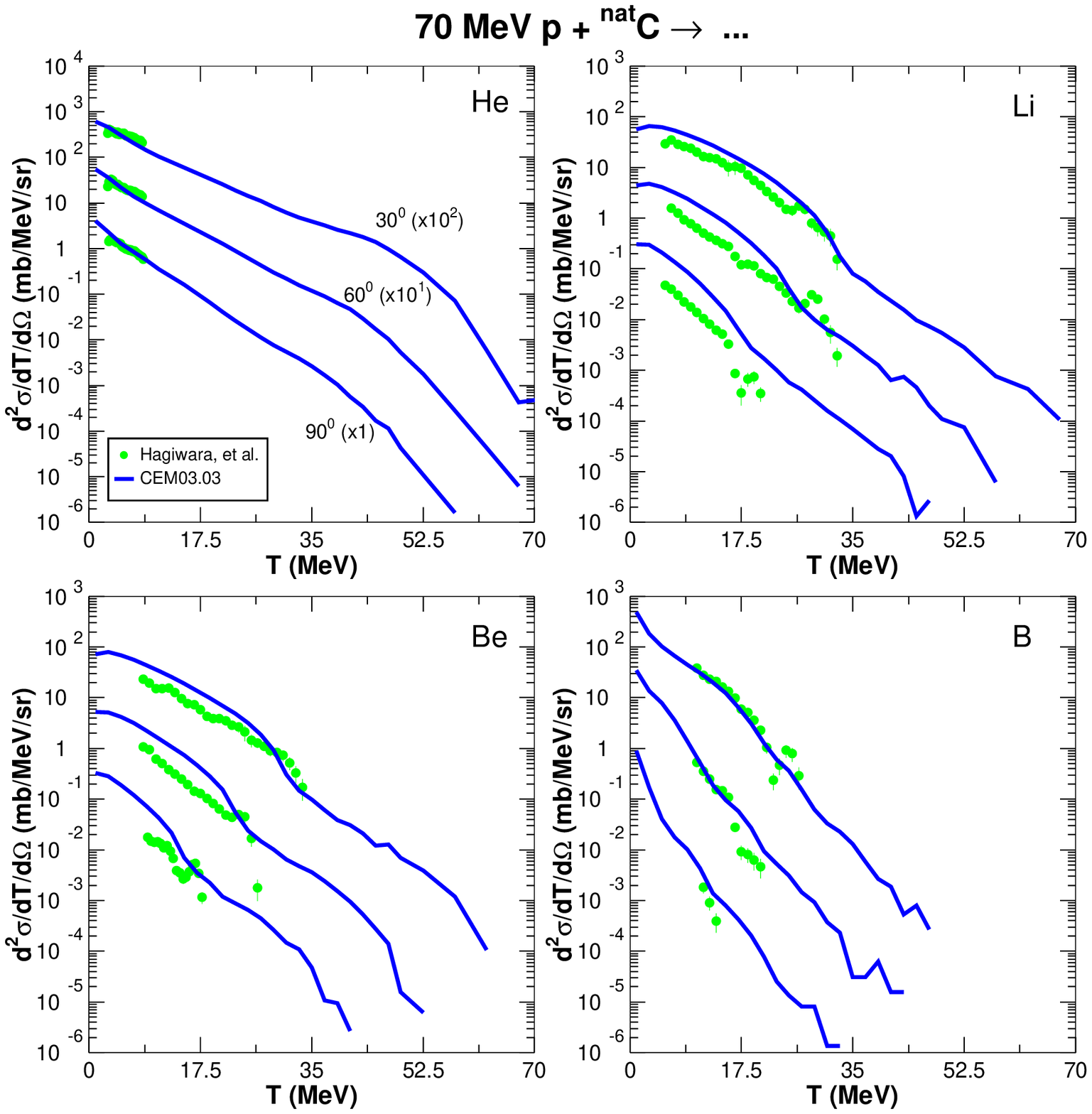}
%\vspace*{2mm}
\caption{Comparison of CEM03.03 (solid lines) He, Li, Be, and B spectra 
from 70 MeV p + C with experimental data by Hagiwara et al. 
\cite{55}
%[55] 
(circles) for a natural carbon target. Our calculations were performed 
for $^{12}$C.
}
\label{fig:24}
\end{figure*}

\clearpage

\begin{figure*}[htb!]
\centering
\includegraphics[width=1.0\textwidth]{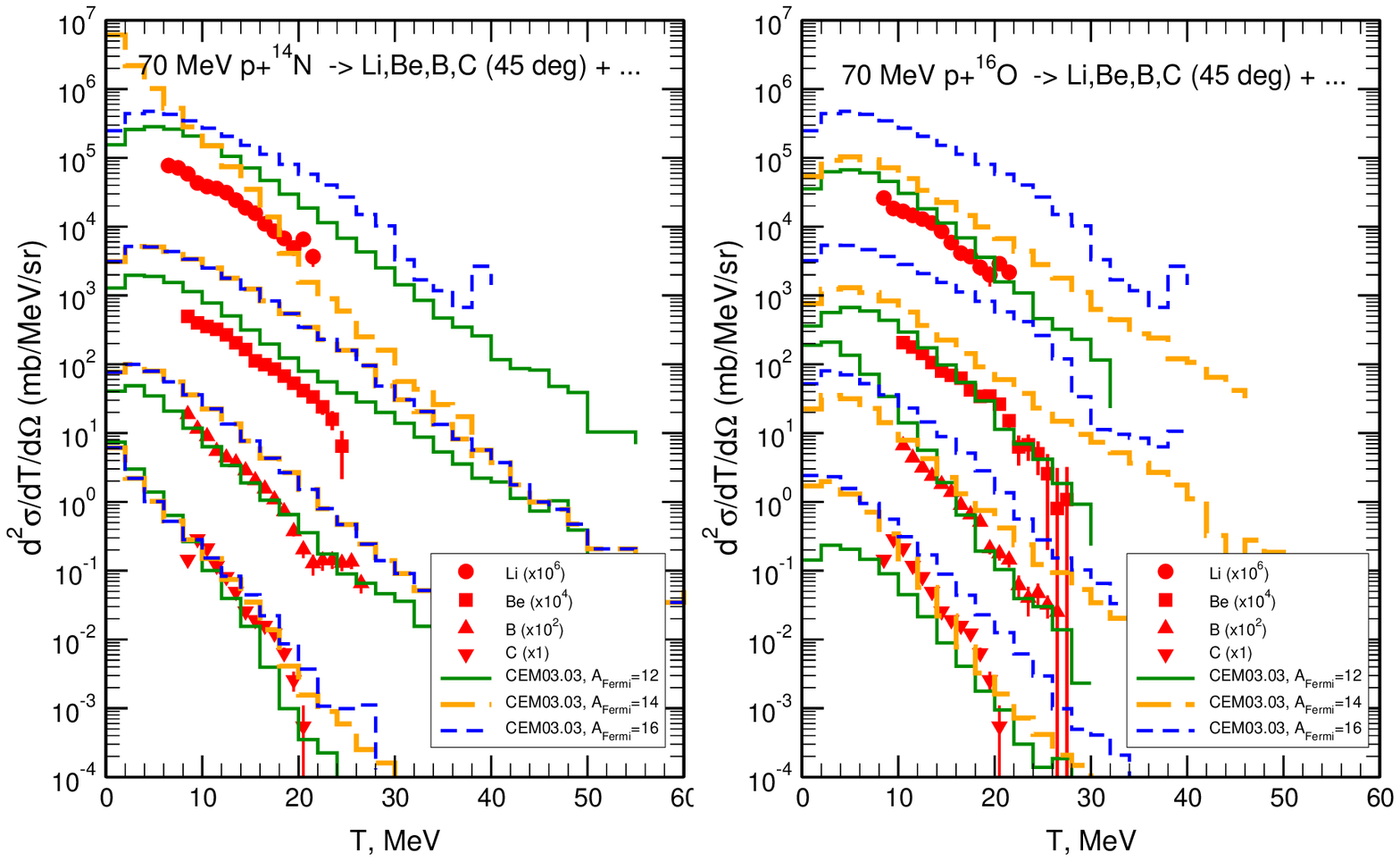}
%\vspace*{2mm}
\caption{Comparison of Li, Be, B, and C spectra at 45 degrees from 70 MeV 
p + $^{14}$N and $^{16}$O measured by Sanami et al. 
\cite{56}
%[56] 
(symbols) with calculations by CEM03.03 using the ``standard'' version of 
the Fermi breakup model ($A_{Fermi} = 12$; solid histograms) and with 
cut-off values for $A_{Fermi}$ of 16 (dashed histograms) and 14 
(long-dashed histograms), as indicated.
}
\label{fig:25}
\end{figure*}

\clearpage

\begin{figure*}[htb!]
\centering
\includegraphics[width=1.0\textwidth]{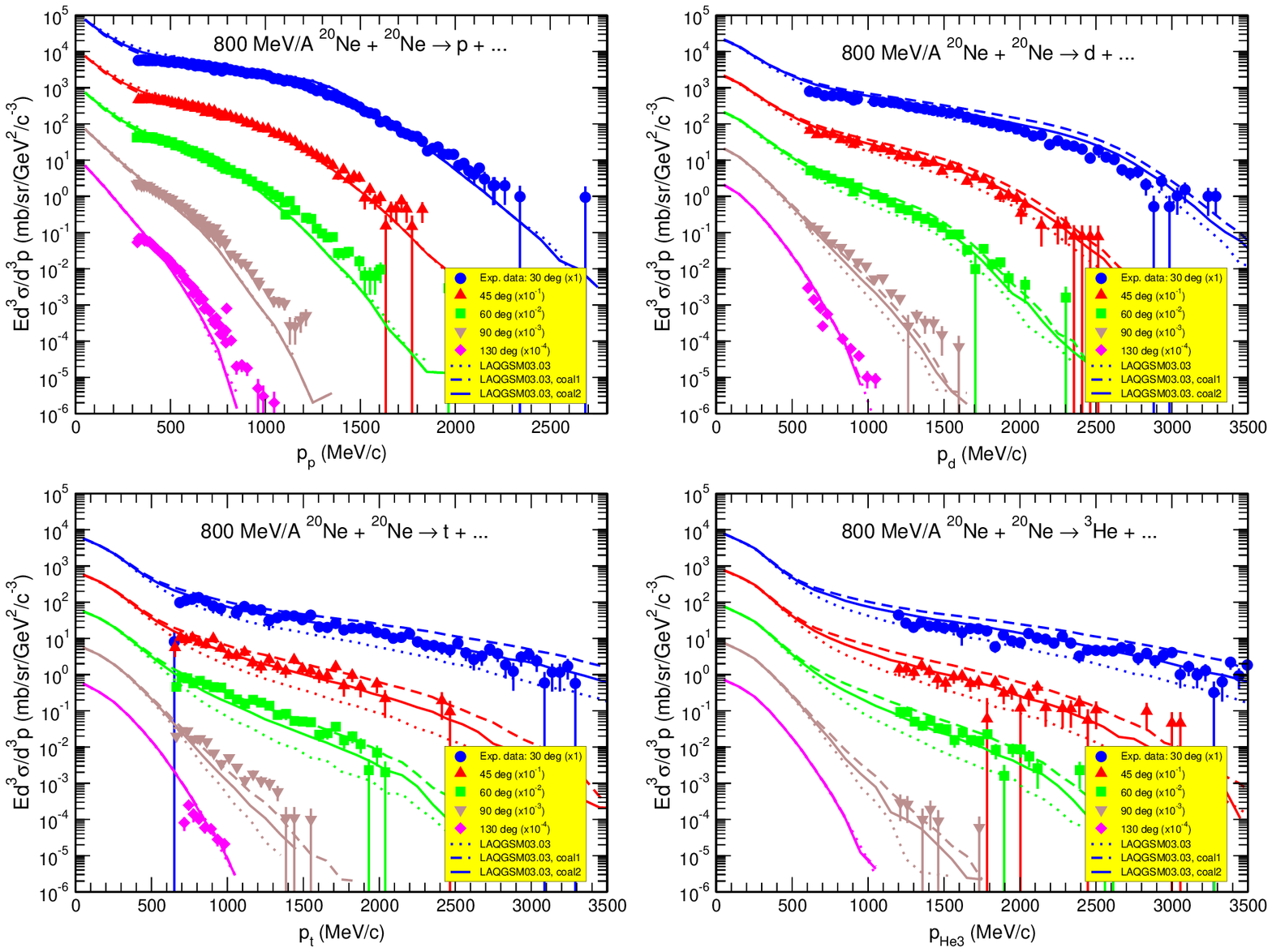}
%\vspace*{2mm}
\caption{Comparison of p, d, t, and $^3$He spectra at 45, 60, 90, and 130 degrees 
from 800 MeV/nucleon $^{20}$Ne + NaF measured at the Bevatron/Bevalac at the 
Lawrence Berkeley Laboratory 
\cite{57, 58}
%[57, 58] 
with calculations by LAQGS03.03 using its ``standard'' version of the coalescence 
model ($p_0 = 0.09$ GeV/c for d, 0.108 GeV/c for t and $^3$He, and 0.115 GeV/c for 
$^4$He; dotted lines) and with modified values of $p_0$ labeled in legend as 
``coal1'' ($p_0 = 0.15$ GeV/c for d, and 0.175 GeV/c for t, $^3$He, and $^4$He; 
dashed lines), as well as with a second modification of $p_0$ labeled in legend as 
``coal2'' ($p_0 = 0.12$ GeV/c for d, and 0.14 GeV/c for t and $^3$He, and $^4$He; 
solid lines), as indicated (for simplicity, all calculations were done on a 
$^{20}$Ne target).
}
\label{fig:26}
\end{figure*}

\clearpage

\begin{figure*}[htb!]
\centering
\includegraphics[width=1.0\textwidth]{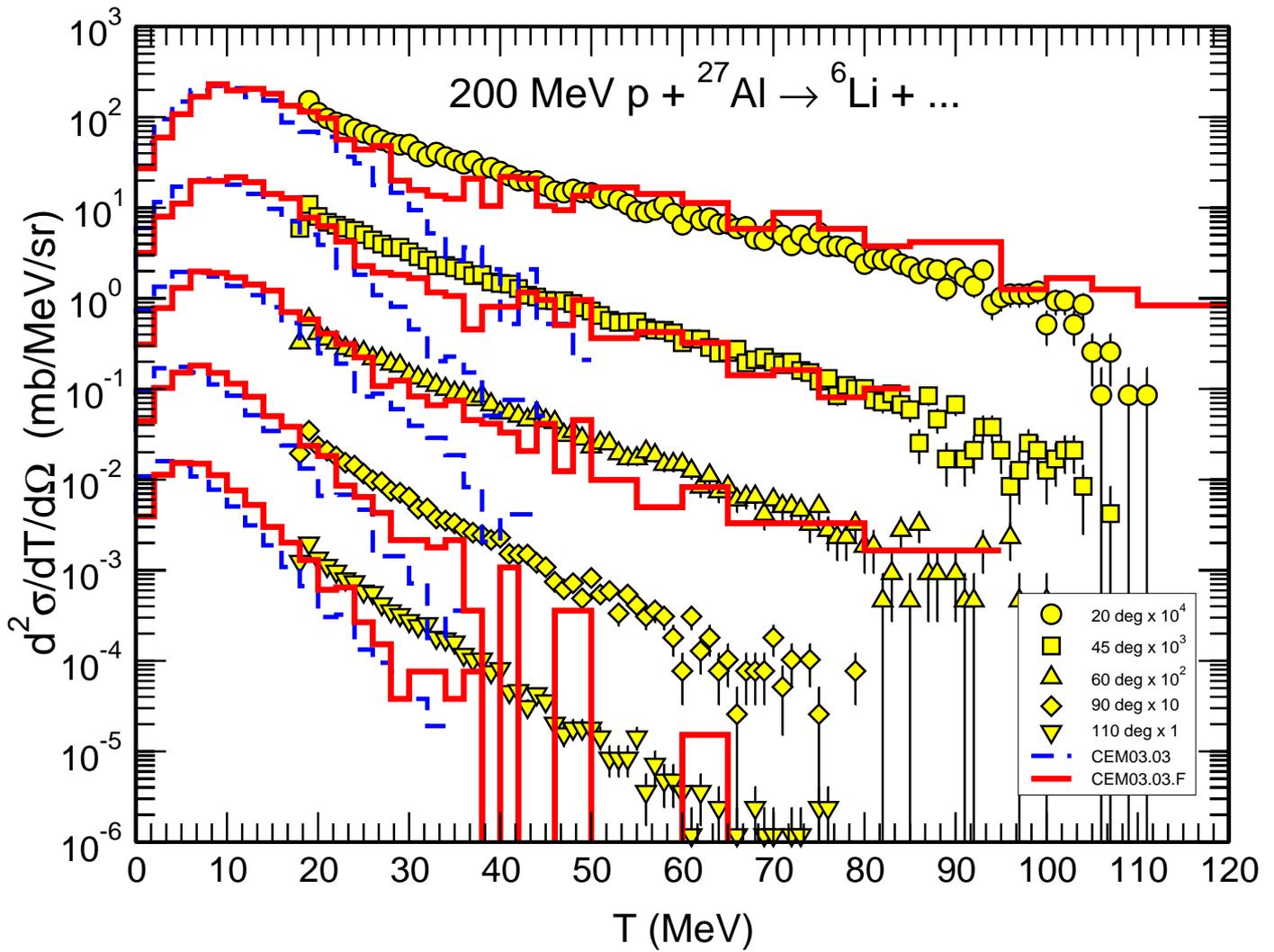}
%\vspace*{2mm}
\caption{Comparison of experimental $^6$Li spectra at 20, 45, 60, 90, and 110 degrees 
by Machner et al. 
\cite{61}
%[61] 
(symbols) with calculations by the unmodified CEM03.03 (dashed 
histograms) and preliminary results with the modified MEM in CEM03.03.F 
(solid histograms), as indicated.
}
\label{fig:27}
\end{figure*}

\end{document}